\documentclass[showpacs, preprintnumbers,
nofootinbib, aps, prd, superscriptaddress,
10pt, showkeys, notitlepage, twocolumn]{revtex4-1}
\usepackage[utf8]{inputenc}
\usepackage{graphicx,amssymb,amsmath,amsthm,amsfonts,epsfig,natbib}
\usepackage[linktocpage]{hyperref}
\usepackage[usenames,dvipsnames]{color}
\usepackage{epstopdf}
\usepackage{float}
\usepackage{amsmath,amssymb,mathrsfs}
\usepackage{tensor}
\usepackage{mathtools}
\usepackage{amsbsy}
\usepackage{bm}
\usepackage{multirow}

\usepackage[scaled]{beramono}

\definecolor{oxfordblue}{rgb}{0.0, 0.13, 0.28}
\definecolor{burgundy}{rgb}{0.5, 0.0, 0.13}
\definecolor{darkolivegreen}{rgb}{0.33, 0.42, 0.18}
\definecolor{darkblue}{rgb}{0,0,0.5}
\definecolor{richcarmine}{rgb}{0.84, 0.0, 0.25}
\definecolor{darkblue}{rgb}{0,0,0.5}
\definecolor{venetianred}{rgb}{0.78, 0.03, 0.08}
\definecolor{skobeloff}{rgb}{0.0, 0.48, 0.45}
\hypersetup{colorlinks=true, citecolor=darkblue, linkcolor=darkblue, urlcolor = darkblue}

\AtBeginDocument{\usepackage{booktabs}}
\newcommand{\ra}[1]{\renewcommand{\arraystretch}{#1}}
\def\be{\begin{equation}}
\def\ee{\end{equation}}
\def\ba{\begin{align}}
\def\ea{\end{align}}

\begin{document}
%--------------------------------------------------------------
\setcounter{topnumber}{1}

\title{Synchronous frequencies of extremal Kerr black holes:\\resonances, scattering and stability}

%--------------------------------------------------------------
\author{Maur\'icio Richartz}
\email{mauricio.richartz@ufabc.edu.br}
\affiliation{Centro de Matem\'atica, Computa\c{c}\~ao e Cogni\c{c}\~ao, Universidade Federal do ABC (UFABC), 09210-170 Santo Andr\'e, S\~ao Paulo, Brazil}
\author{Carlos A.~R.~Herdeiro}
\email{herdeiro@ua.pt}
\affiliation{Departamento de F\'isica da Universidade de Aveiro and Center for Research and Development in Mathematics and Applications (CIDMA), Campus de Santiago, 3810-183 Aveiro, Portugal}
\author{Emanuele Berti}
\email{eberti@olemiss.edu}
\affiliation{Department of Physics and Astronomy, The University of Mississippi,
University, MS 38677-1848, USA}
\affiliation{CENTRA, Departamento de F\'{\i}sica, Instituto Superior
T\'ecnico, Universidade de Lisboa,
Avenida Rovisco Pais 1, 1049 Lisboa, Portugal}

%--------------------------------------------------------------
%------------------------------------------------------------------------------------------------------------------------------------------
%------------------------------------------------------------------------------------------------------------------------------------------
%------------------------------------------------------------------------------------------------------------------------------------------
%------------------------------------------------------------------------------------------------------------------------------------------
 
%------------------------------------------------------------------------------------------------------------------------------------------

%--------------------------------------------------------------
\begin{abstract}
%--------------------------------------------------------------

The characteristic damping times of the natural oscillations of a Kerr black hole become arbitrarily large as the extremal limit is approached. This behavior is associated with the so-called zero damped modes (ZDMs), and suggests that extremal black holes are characterized by quasinormal modes whose frequencies are purely real. Since these frequencies correspond to oscillations whose angular phase velocity matches the  horizon angular velocity of the black hole, they are sometimes called ``synchronous frequencies''. Several authors have studied the ZDMs for near-extremal black holes. Recently, their correspondence to branch points of the Green's function of the wave equation was linked to the Aretakis instability of extremal black holes. Here we investigate the existence of ZDMs for extremal black holes, showing that these real-axis resonances of the field are unphysical as natural black hole oscillations: the corresponding frequency is always associated with a scattering mode. By analyzing the behavior of these modes near the event horizon we obtain new insight into the transition to extremality, including a simple way to understand the Aretakis instability. 

%--------------------------------------------------------------
\end{abstract}
%--------------------------------------------------------------
\maketitle
%
%
%================
\section{Introduction}
%================

The gravitational wave signal produced by a perturbed rotating (Kerr) black hole can be roughly divided into three phases~\cite{Kokkotas:1999bd,Nollert:1999ji,Berti:2009kk,Konoplya:2011qq}: the initial burst of radiation, the ringdown phase, and the late-time tail. The ringdown phase is dominated by damped oscillations of the form $e^{-i \omega t}$ with complex quasinormal mode (QNM) frequencies $\omega$. The real part of $\omega$ is the oscillation frequency, while the imaginary part is associated with the characteristic damping time. A remarkable fact is that the quasinormal frequency spectrum depends only on the black hole parameters, namely its mass $M$ and its specific angular momentum $a$ (throughout this paper we assume $G=c=1$).

 Theoretically, the angular momentum of a Kerr black hole is bound from above by $a=M$, in which case it is said to be extremal.  The angular velocity of the black hole depends on the ratio $a/M$, and is denoted by $\Omega_h$ [for an extremal black hole, $\Omega_h=(2M)^{-1}$]. In astrophysical scenarios, it is believed that black holes cannot rotate faster than $a \approx 0.998M$. This is the Thorne limit~\cite{1974ApJ...191..507T}, predicted from the analysis of a general accretion process by a black hole; see however~\cite{Sadowski:2011ka} (in particular Section 1.1 and references therein), which predicts a limit closer to extremality.

Despite this limitation, it is nonetheless interesting and important to study extremal black holes, at least from a theoretical perspective. In fact, extremal black holes are characterized by vanishing Hawking temperatures and therefore, even semiclassicaly, do not radiate. Hence, they are believed to possess a simpler description (compared to nonextremal ones) in a complete quantum gravity theory~\cite{Strominger:1996sh,Guica:2008mu}.

Since Kerr black holes are widely used in both theoretical analyses and astrophysical applications~\cite{Wiltshire:2009zza}, it is crucial to discuss their stability. Mode stability (the absence of characteristic oscillations with definite frequency that grow in time) was verified by Whiting for nonextremal black holes in~\cite{Whiting:1988vc} (see also~\cite{1973ApJ...185..649P,1973ApJ...185..675D,1974PhRvL..32..243F}), and for extremal black holes in~\cite{Richartz:2015saa}. Linear stability, on the other hand, concerns the existence of perturbations of generic initial data that grow in time. For nonextremal black holes, linear stability was rigorously demonstrated only very recently~\cite{Dafermos:2010hd,Dafermos:2014cua}, mainly due to the difficulty of proving that no quasinormal frequencies exist on the real axis (this was accomplished in \cite{Shlapentokh-Rothman:2013hza}). 

Aretakis proved the remarkable result that extremal black holes are linearly unstable~\cite{Aretakis:2011gz,Aretakis:2011ha,Aretakis:2011hc,Aretakis:2012bm} (see also~\cite{Lucietti:2012sf,Lucietti:2012xr,Murata:2013daa}).
The polynomial (rather than exponential) nature of the Aretakis instability raised the concern that it could not be explained through a mode analysis. 
Such concerns have been recently laid to rest by Casals {\it et al.}~\cite{Casals:2016mel}, who showed that the instability can be understood as branch points (located at $\omega = m \Omega_h$  for every $m \in \mathbb{Z}$) of the frequency domain Green's function ~\cite{1974CMaPh..38...47H,Glampedakis:2001js}.

The explanation of these branch points can be traced back to the work of Detweiler~\cite{1980ApJ...239..292D}, who used the analytical results of Press and Teukolsky~\cite{1974ApJ...193..443T} to prove the existence of very long-lived quasinormal oscillations for near-extremal black holes, nowadays called zero-damped modes (ZDMs). In fact, as the extremal limit is approached, an infinite number of QNM overtones appears to converge to each frequency $m\Omega_h$, with $m \in \mathbb{Z}^*$ (note that we do not fix $\mathrm{Re}(\omega) > 0$, meaning that ZDMs exist for both positive and negative $m$). 
This result naturally leads to the conjecture that extremal black holes have infinitely long-lived quasinormal oscillations with real frequencies given by $\omega = m \Omega_h = m / (2M)$. For this reason, Detweiler considered extremal black holes marginally stable. After Detweiler, several investigators studied ZDMs of near-extremal black holes, both analytically and numerically~\cite{Detweiler:1983zz,Ferrari:1984zz,Sasaki:1989ca,Glampedakis:2001js,Andersson:1999wj,Onozawa:1996ux,Cardoso:2004hh,Hod:2008zz,Yang:2012pj,Yang:2013uba, Nakano:2016zvv}.

The Aretakis instability has been already elucidated from the point of view of mode analysis~\cite{Casals:2016mel} 
(see also \cite{Gralla:2016sxp,Zimmerman:2016qtn}), but it is interesting to investigate whether ZDMs correspond to solutions of the wave equation in the frequency domain for extremal black holes. Ref.~\cite{Casals:2016mel} puts forward an argument based on energy conservation as to why ZDMs should not exist in the extremal limit. The primary goal of this work is to investigate in detail whether ZDMs exist for extremal black holes, and the physical nature of the branch points. Instead of relying on the limit $\omega \rightarrow m \Omega_h$, as typically done in almost every related analysis, we will study what happens exactly at $\omega = m \Omega_h$ for an extremal black hole. 

We will show that the wave character of the mode solutions is lost near the event horizon for ZDMs. We establish in Sec.~IV that the corresponding frequency is never a quasinormal (or normal) frequency. It is, instead, associated with \textit{scattering modes} if the condition
\be \label{reg_condition}
\alpha_{s \ell m} \equiv \lambda_{s\ell m}\left(\frac{m}{2}\right) - \frac{7}{4}m^2 > 0,
\ee
which ensures regularity at the horizon, is satisfied. Here $s=0,\pm 1,\pm 2$ is the spin-weight, $m$ is the azimuthal wave number, $\ell$ is the orbital wave number, and $\lambda_{s\ell m}(m/2)$ is the spin-$s$ spheroidal eigenvalue~\cite{1973ApJ...185..635T,Berti:2005gp} $\lambda_{s\ell m}(a\omega)$ evaluated at $a\omega=m/2$. 
For some values of the parameters $(s,\ell, m)$, as shown in Figs.~\ref{fig1} and \ref{fig2} below,
the condition (\ref{reg_condition}) does not hold and mode solutions are never regular at the event horizon. Even when condition~\eqref{reg_condition} holds, sufficiently high-order radial derivative of regular modes have an irregular behaviour on the horizon. This suggests an intuitive way to picture the Aretakis instability as being triggered by synchronous modes scattering off an extremal Kerr black hole. In the rest of the paper we provide details in support of these conclusions.

%================
\section{Wave equation, asymptotic solutions, and zero-damped modes}
%================

    A revolution in the field of black hole perturbations was initiated in the early 1970s. Working in the Newman-Penrose (NP) formalism~\cite{1962JMP.....3..566N}, and using the Kinnersley tetrad~\cite{1969JMP....10.1195K}, Teukolsky~\cite{Teukolsky:1972my,1973ApJ...185..635T} showed that the perturbation equations are separable when one defines appropriate scalar functions $\Upsilon_s$.  The simplest case is that of a scalar perturbation, described by a massless Klein-Gordon field represented by the single function $\Upsilon_0$. Electromagnetic perturbations, on the other hand, are described by the functions $\Upsilon_{-1}$ and $\Upsilon_{1}$, which relate, respectively, to the NP electromagnetic scalars $\phi_0$ and $\phi_2$. Gravitational perturbations are similarly described by the functions $\Upsilon_{-2}$ and $\Upsilon_{2}$, related to the Weyl scalars $\psi_0$ and $\psi_4$, respectively.

In the usual Boyer-Lindquist coordinates~\cite{Boyer:1966qh} $(t,r,\theta,\phi)$, the Teukolsky fields can be separated by writing $ \Upsilon_s = R_{s \ell m \omega}(r)S_{s \ell m }(\theta, a\omega) e^{i m \phi - i \omega t}$. 
To simplify notation, we will write $\Upsilon_s = R_s(r)S_s(\theta) e^{i m \phi - i \omega t}$. The angular functions $S_s (\theta) = S_{s \ell m}(\theta,a\omega)$ are the (frequency-dependent) spin-weighted spheroidal harmonics~\cite{1973ApJ...185..635T,Berti:2005gp}. The radial wave function $R_s(r)$ obeys the so-called radial Teukolsky equation,
\begin{align}
\nonumber  \Delta \frac{d^2 R_s}{dr^2} + 2(s+1)(r-M)  \frac{dR_s }{dr}   + \\
\left(
\frac{K^2 -  2is (r-M)K}{\Delta} + 4is\omega r +  X_s \right)R_s   =0, \label{teukorad}
\end{align}
where $\Delta = r^2 - 2Mr + a^2$, $K=\omega (r^2 + a^2) -am$, $ X_s = 2 m a \omega - a^2 \omega^2 - \lambda_{s\ell m}(a\omega)$, and $\lambda_{s\ell m}(a\omega)$ is a separation constant.   

For a nonextremal black hole, the smallest and the largest roots of the function $\Delta$, namely $r_-=M-\sqrt{M^2 - a^2}$ and $r_+=M+\sqrt{M^2 - a^2}$, determine, respectively, the locations of the Cauchy horizon and of the event horizon of the black hole. In the extremal case, the two horizons coincide at $r=r_+=r_-=M$. Invariance under rotations $\phi \rightarrow \phi + 2\pi$ implies that $m \in \mathbb{Z}$. By requiring the angular function to be regular at $\theta = 0$ and at $\theta = \pi$, only a discrete set of separation constants $\lambda_{s \ell m}(a\omega)$, where $\ell \ge \max(|m|,|s|)$ is a positive integer, is allowed.

The most general solution of the radial Teukolsky equation can only be written in terms of confluent Heun functions~\cite{1983NCimL..37..300M,1983NCimL..38..561B,Fiziev:2009wn,Cook:2014cta}, but its asymptotic solutions have a simple analytic form. In terms of the tortoise coordinate $r^*$, defined by $dr^*/dr =  (r^2 + a^2)/\Delta$, and the function $Y_s(r)= \Delta^{s/2} \sqrt{r^2+a^2} R_{s}(r)$, we have
\be \label{near1}
Y_s(r^*) = K_{\rm in}^{\rm h} (r-r_+)^{-s/2}e^{-i k r^*} + K_{\rm out}^{\rm h} (r-r_+)^{s/2} e^{i k r^*}
\ee
near the event horizon ($r^* \rightarrow -\infty$, $r\rightarrow r_+$), and
\be \label{far1}
Y_s(r^*) = K_{\rm in}^\infty r^{s}e^{-i\omega r^*} + K_{\rm out}^\infty r^{-s} e^{i \omega r^*}
\ee
far away from the black hole ($r^* \rightarrow \infty$, $r\rightarrow \infty$).
Here $K_{\rm in}^{\rm h}$, $K_{\rm out}^{\rm h}$, $K_{\rm in}^\infty$ and $K_{\rm out}^\infty$ are constants, $r=r(r^*)$, and $k = \omega - m \Omega_h$, with $\Omega_h = a/(2Mr_+)$, is the effective wavenumber near the horizon.

The simplest way to distinguish between ingoing and outgoing waves is to calculate the group velocity $\partial \omega / \partial k$ for each mode solution. Recognizing that the dispersion relation is $k^2=\omega ^2$ far away, and $k^2 = (\omega - m \Omega_h)^2$ near the event horizon, it is straightforward to conclude that the waves associated with $K_{\rm in}^{\rm h}$ and $K_{\rm in}^\infty$ are ingoing, while the ones associated with $K_{\rm out}^{\rm h}$ and $K_{\rm out}^\infty$ are outgoing. The fact that, classically, nothing can escape from the black hole translates into the boundary condition $K_{\rm out}^{\rm h} = 0$. When studying wave scattering processes, one fixes the amplitude $K_{\rm in}^\infty \neq 0$ of the incident wave, and then solves the radial Teukolsky equation to determine $K_{\rm in}^{\rm h}$ and $K_{\rm out}^\infty$ for a given frequency $\omega$.

 A peculiar behavior occurs when $0<\omega < m \Omega_h$: the radial phase and group velocities have different signs, meaning that the ingoing wave will appear to be outgoing to an asymptotic observer. Since the energy flux is associated with the phase velocity of the wave, this fact is usually invoked to explain the phenomenon of superradiance~\cite{1971JETPL..14..180Z,1972JETP...35.1085Z,1972BAPS...17..472M,Starobinsky:1973aij,1974JETP...38....1S}, in which rotational energy is extracted from the black hole (see also~\cite{Bekenstein:1998nt,Richartz:2009mi,Brito:2015oca} for more recent accounts on the topic, and \cite{Torres:2016iee} for the first experimental observation of the effect). On the contrary, when studying resonances of the system, the natural boundary condition far away from the black hole is that no incident waves exist, i.e.~$K_{\rm in}^\infty=0$. In such a case, only a countably infinite set of frequencies (the QNM frequencies) can satisfy both boundary conditions simultaneously.   

This is the standard description for generic perturbations and scattering processes around a nonextremal rotating black hole. It is also standard procedure to associate a Wronskian with the equation for $Y_s(r^*)$. The constancy of the Wronskian implies the relation
\be \label{reflec0}
\mathcal{R}_s = 1 - \frac{\omega - m\Omega_h}{\omega} \mathcal{T}_s,
\ee
where the reflection coefficient $\mathcal{R}_s$ is defined as the ratio between the ingoing and the outgoing energy fluxes at infinity, and the transmission coefficient $\mathcal{T}_s$ is defined as the ratio between the ingoing energy fluxes at the event horizon and at infinity. Using the stress energy tensor, it is possible to express $\mathcal{R}_s$ and $\mathcal{T}_s$ in terms of $K_{\rm in}^{\rm h}$, $K_{\rm in}^\infty$, and $K_{\rm out}^\infty$.
Note, however, that for the particular frequencies $\omega = 0$ and $\omega = m \Omega_h$ the picture above is incomplete. If either $\omega = 0$ or $\omega = m \Omega_h$, Eqs.~\eqref{near1} and \eqref{far1} are not the most general asymptotic behaviors, because the two terms appearing in each case become linearly dependent. A more detailed investigation is desirable at these frequencies. 

This consideration led Hod to prove that massive scalar fields allow the existence of stationary bound states around Kerr black holes~\cite{Hod:2012px}. These bound states, called stationary scalar clouds, occur only when $\omega = m\Omega_h$. They were later considered by Herdeiro and Radu~\cite{Herdeiro:2014goa} in a fully nonlinear Einstein-Klein-Gordon theory (taking into account the field's backreaction), leading to the discovery of black holes with scalar hair~\cite{Herdeiro:2015gia}. For $\omega = m\Omega_h$ the waves become synchronized with the black hole, since their angular phase velocity $\omega / m$ matches the horizon angular velocity. Therefore we shall refer to $m\Omega_h$ (which is also the critical superradiant frequency) as the \textit{synchronous frequency}.      

The ZDMs are intimately related to these synchronous frequencies. Starting with the wave equation for generic perturbations around a nonextremal Kerr black hole, the standard procedure~\cite{Starobinsky:1973aij,1974JETP...38....1S,1974ApJ...193..443T} is to take the double limit ($a\rightarrow M$, $\omega \rightarrow m\Omega_h$), where $m\in \mathbb{Z}$, resulting in QNMs whose frequencies $\omega$ are such that $\mathrm{Re}\left( \omega \right) = m \Omega_h + \mathcal{O}\left( M^2 - a^2 \right)$ and $\mathrm{Im}\left( \omega \right) = \mathcal{O}\left( \sqrt{M^2 - a^2} \right)$.
Assuming that each quasinormal frequency is a continuous function of the spin parameter $a$, one can extrapolate the results above to conclude that the corresponding QNMs for extremal black holes are purely real, i.e.~$\omega _{\rm extremal} = \lim_{a \rightarrow M} \omega \left(a \right) = m \Omega_h$.

 While these double limit calculations are valid for near-extremal black holes, one should be careful when extrapolating them to the extremal case by continuity arguments. First of all, unlike the nonextremal case, the frequencies of these purely real modes for extremal black holes would be exactly the synchronous frequencies. We have just shown that, in this case, the standard treatment of the boundary conditions \eqref{near1} is inappropriate, meaning that the extremal and nonextremal cases of the ZDMs have to be analyzed separately. Secondly, for a generic frequency and a nonextremal black hole, the event horizon is a regular singular point of the radial Teukolsky equation, while for an extremal black hole, it is an irregular singularity, leading to an exponentially singular wave behavior~\cite{Richartz:2015saa}. Furthermore, if the synchronous frequency is assumed, the nature of the horizon changes again. As we will see, it becomes a regular singular point with logarithmic solutions for nonextremal black holes, and a regular singular point without logarithmic solutions for extremal ones.

The arguments above suggest that any result for extremal black holes derived from continuity arguments applied to near-extremal black holes must be treated with great care. In fact, originally motivated by calculations indicating that the entropy of extremal black holes vanishes~\cite{Hawking:1994ii,Teitelboim:1994az}, several authors argued that the transition from a nonextremal to an extremal black hole is not continuous~\cite{Das:1996rn, Carroll:2009maa, Pradhan:2010ws, Pradhan:2012yx}. For instance, whether near-extremal geometries are asymptotically flat or not depends on the limiting procedure as $a \rightarrow M$~\cite{Gralla:2015rpa}. Similarly, results for the synchronous frequency (and for zero frequency) based on limits of the corresponding results for generic frequencies must be interpreted with care. In particular, the transition of the ZDMs of near-extremal black holes to hypothetical ZDMs of extremal black holes must be studied carefully. In other words, for a given overtone, the quasinormal frequency of the ZDMs might not be a continuous function of the spin parameter $a$ at $a=M$ after all, invalidating the previous continuity argument in favor of their existence.

\begin{table*}\centering
\ra{2}
\begin{tabular}{@{}llllll@{}}\toprule
Synchronous &  Extremal & $R_s^I(r)$ & $R_s^{II}(r)$ & $G_s^I(r)$ & $G_s^{II}(r)$ \\ \midrule
 No & No & $(r-r_+)^{-s-iJ_0}$                                                  & $(r-r_+)^{iJ_0}$                                                                 & 1 & $(r-r_+)^{-s+2iJ_0}$\\
 No & Yes & $e^{\frac{iJ_1}{r-M}}(r-M)^{-2s - 2iM\omega}$ & $e^{\frac{-iJ_1}{r-M}}(r-M)^{2iM\omega}$          & $1$ & $e^{\frac{-2iJ_1}{r-M}}(r-M)^{-2s+4iM\omega}$ \\ 
  \multirow{2}{*}{Yes}  & \multirow{2}{*}{No} &  \multirow{2}{*}{$(r-r_+)^{\xi_I}$} & $(r-r_+)^{\xi_{II}}$, \thinspace  if $s\neq 0$ & \multirow{2}{*}{$(r-r_+)^{\xi_I}$} & $(r-r_+)^{\xi_{II}}$,\ if $s\neq 0$ \\  &&&$\log (r-r_+)$, if $s=0$&&$\log (r-r_+)$, if $s=0$ \\
 Yes & Yes & $(r-M)^{-\frac{1}{2} - s + \delta_s}$ & $(r-M)^{-\frac{1}{2} -s - \delta_s}  $ & $ (r-M)^{-\frac{1}{2} +im - s + \delta_s}$ & $(r-M)^{-\frac{1}{2} +im -s -\delta_s }  $\\
\bottomrule
\end{tabular}
\caption{The different near-horizon behaviors of the field for the two radial functions $R_s(r)$ and $G_s(r)$, defined as $\Upsilon_s = R_s(r)S_s(\theta) e^{i m  \phi - i \omega t}$ and $\Gamma_s = G_{s}(r)S_{-s}(\theta) e^{i m \tilde \phi - i \omega v}$. The parameters are $J_0= 2Mr_+(\omega - m \Omega_h) / ( r_+ - r_ -)$, $J_1= 2M^2(\omega - m \Omega_h)$, $\xi_I=\max\{0,-s\}$, $\xi_{II}=\min\{0,-s\}$, and $\delta_s = \sqrt{\left(\frac{1}{2} + s\right)^2 - \frac{7}{4}m^2 + \lambda_{s\ell m}\left(\frac{m}{2}\right)}$. Logarithmic terms  are present in the synchronous, nonextremal cases also when $s \neq 0$, but they are never of leading order near the horizon.} \label{table}
\end{table*}

 \section{Near-horizon behavior and boundary conditions for synchronous frequencies} \label{sec3}

Let us now see explicitly how the near-horizon behavior changes when the synchronous frequency $\omega = m \Omega_h$ is assumed from the beginning. In this case, for both nonextremal and extremal black holes, the event horizon is a regular singular point of the wave equation. Hence, one can use the Frobenius method to find power series solutions around the event horizon. In fact, the two independent solutions for a nonextremal black hole are 
\be 
R^{I}_{s}(r) \sim (r-r_+)^{\xi_I}, \label{log1a} 
\ee
and
\be
R^{II}_{s}(r) \sim Z_s \log(r-r_+) R^{I}_{s}(r) +  (r-r_+)^{\xi_{II}}, \label{log1b}
\ee
plus higher order corrections, where $\xi_I=\max\{0,-s\}$, $\xi_{II}=\min\{0,-s\}$, and $Z_s \neq 0$ is a different constant for each $s$. Logarithmic terms are always present, but the leading-order contribution of $R_s^{II}$ is logarithmic only when $s=0$.

For an extremal black hole, on the other hand, we have
\be \label{near5}
R_{s}^{I,II} \sim (r-M)^{-\frac{1}{2}-s\pm \delta_s}
\ee
plus next-to-leading order corrections, where
\be \label{deltas}
\delta_s^2 = \delta_{s}^2(\ell,m) = \left(\frac{1}{2} + s\right)^2 - \frac{7}{4}m^2 + \lambda_{s \ell m}\left(\frac{m}{2}\right),
\ee
and the upper (lower) sign refers to the index I (II). The equation above defines $\delta_s$ up to a sign, so without loss of generality we assume $\delta_s=\sqrt{\delta_s^2}$, i.e.~$\mathrm{Re}(\delta_s) > 0$ in general and $\mathrm{Im}(\delta_s) > 0$ when its real part vanishes. Since $\lambda_{-s \ell m}(m/2)=\lambda_{s \ell m}(m/2) + 2s$~\cite{Berti:2005gp}, the parameter $\delta_s$ is independent of the sign of $s$. Additionally, since the argument $a\omega=m/2$ of $\lambda_{s \ell m}(a\omega)$ is real, the separation constant $\lambda_{s \ell m}(m/2)$ is itself real and, consequently, $\delta_s$ is either real or purely imaginary.

The typical oscillatory behavior is lost here (even in the tortoise coordinates). Fortunately, in order to determine the natural boundary conditions at the event horizon, there is a more general procedure than the one described in the previous section~\cite{1972ApJ...178..347B,1973ApJ...185..635T}. It consists in demanding that the wave function be regular at the event horizon, so that the energy-momentum tensor is well behaved and the test field approximation holds. The Boyer-Lindquist coordinates, being singular at the horizon, are ill suited for this purpose. We therefore need to resort to regular coordinates at the horizon. One such example, that we shall use here, corresponds to the ingoing Kerr coordinates $(v,r,\theta,\tilde \phi)$~\cite{1973ApJ...185..635T}, related to Boyer-Lindquist coordinates $(t,r,\theta,\phi)$ by
\be
dv = dt + \frac{r^2 + a^2}{\Delta}dr, \quad d \tilde \phi  = d \phi + \frac{a}{\Delta}dr. 
\ee

The Kinnersley tetrad is not well behaved at the future event horizon. We use instead the tetrad proposed by Hartle and Hawking~\cite{1972CMaPh..27..283H,1974ApJ...193..443T}, which is obtained from the Kinnersley tetrad by the transformation $(t \rightarrow -t, \varphi \rightarrow -\varphi)$. With this new tetrad, the Teukolsky fields $\Upsilon_s$ change to new functions $\Gamma_s = 2^s \Delta^{-s} \Upsilon_{-s}$. Separating these new fields in ingoing Kerr coordinates as $\Gamma_s=G_s(r)S_{-s}(\theta)e^{-i\omega v}e^{i m \tilde \phi}$, Teukolsky and Press~\cite{1974ApJ...193..443T} were able to show that the new radial function $G_s(r)$ satisfies
\begin{align}
\nonumber  \Delta \frac{d^2 G_s}{dr^2} + \left[2(s+1)(r-M) - 2 i K\right]  \frac{dG_s}{dr}   + \\
\left[
-2(2s+1)i\omega r + X_s \right]G_s   =0, \label{teukorad2}
\end{align}
where the functions $K$ and $X_s$ are the same as in Eq.~\eqref{teukorad}. Alternatively, using the relations between $\Gamma_s$ and $\Upsilon_s$ and between the coordinates $(v,\tilde \phi)$ and $(t, \phi)$, we can show that the relation between $G_s$ and $R_s$ is
\begin{align}
G_{s}(r) = R_{-s}(r) 2^s (r-r_+)^ {-s} (r-r_-)^{-s+2iM\omega} \nonumber   \\
 \times \exp \left(i \omega r + 2iMr_+\left(\omega - m \Omega_h \right) \int_r \frac{dr'}{\Delta}\right). 
\end{align}

In terms of $G_s$, 
the near-horizon behavior of a spin-$s$ field with the synchronous frequency coincides with Eqs.~\eqref{log1a} and \eqref{log1b} for nonextremal black holes, the only difference being the constant $Z_s$. 
On the other hand, the solution for extremal black holes changes slightly to
\begin{align} \label{GIII}
G^{I,II}_{s}(r) \sim (r-M)^{-\frac{1}{2} +im - s \pm \delta_s},    
\end{align}
plus next-to-leading-order terms.
For comparison and completeness, all possible variations of frequency (i.e.~synchronous or not) and black hole spin (i.e.~extremal or not), with their respective solutions, are shown in Table \ref{table}. 

Since the Hartle-Hawking tetrad in ingoing Kerr coordinates is well-behaved for a physical observer located at the future event horizon, it is natural to take as a boundary condition that $\Gamma_s$ (and therefore $G_s$) must be regular at the event horizon. For a generic frequency $\omega \neq m \Omega_h$, this prescription is equivalent to imposing the boundary condition of an ingoing group velocity at the event horizon. For synchronous frequencies this is the only possibility, since the wave character of the solutions is lost. From the appropriate boundary condition for $G_s(r)$, it is straightforward to use the relations above to determine the corresponding boundary condition for $R_s(r)$.

In terms of the two independent solutions, the most general solution of the wave equation can be written as 
\be
G_s(r) = C^I_s G_s^{I} (r) + C^{II}_s G_s^{II} (r), 
\ee
where $C^I_s$ and $C^{II}_s$ are constants (of course, a similar relation holds for $R_s$ in terms of $R_s^{I}$ and $R_s^{II}$). Note that, even though the Teukolsky equations for $+s$ and $-s$ separate, they provide the same physical information~\cite{1973ApJ...185..649P}, meaning that one can never have $\Gamma_s$ without $\Gamma_{-s}$ (or, equivalently, $\Upsilon_s$ without $\Upsilon_{-s}$). This means that $G_s$ and $G_{-s}$ are not independent. In fact, they are related by the Teukolsky-Starobinski identities~\cite{1974ApJ...193..443T} (usually given in terms of $R_s$ instead of $G_s$):
\be 
\mathcal{D}^2 R_{-1} = \frac{B_{\rm em}}{2} R_1, \quad \left(\mathcal{D^{\dagger}}\right)^2 \Delta R_{1} = \frac{2B_{\rm em}^*}{\Delta} R_{-1}, 
\ee
for electromagnetic perturbations, and 
\be 
\mathcal{D}^4 R_{-2} = \frac{1}{4}B_{\rm grav} R_2, \quad \left(\mathcal{D^{\dagger}}\right)^4 \Delta^2 R_{2} = \frac{4B_{\rm grav}^*}{\Delta^{2}} R_{-2}, 
\ee
for gravitational perturbations, where $\mathcal D = \partial_r - i K/ \Delta$ and $\mathcal D^\dagger = \partial_r + i K / \Delta$ are differential operators, and $B_{\rm em}$, $B_{\rm grav}$ are the so-called Starobinski-Churilov constants.
A direct consequence of these identities, after plugging in the formulas for $R_s^{I}$ and $R_s^{II}$ given in Table~\ref{table}, is the fact that $C^{I}_s$ and $C^{II}_s$ are proportional, respectively, to $C^{I}_{-s}$ and $C^{II}_{-s}$, i.e.~
\be \label{srelation}
C^{I}_s \propto C^{I}_{-s}, \qquad C^{II}_s \propto C^{II}_{-s}.
\ee

We can now formulate the appropriate boundary condition at the event horizon in terms of $C^{I}_s$ and $C^{II}_s$. Notice first that, for the first three cases in Table \ref{table}, $G_s^I(r)$ and all its radial derivatives are well-behaved at the event horizon, while $G_s^{II}(r)$ is always divergent for $s \geq 0$. Hence, the regularity requirement for $\Gamma_s$ translates into the boundary condition $C^{II}_s = 0$ for $s \geq 0$. By virtue of the Teukolsky-Starobinski identities, through \eqref{srelation}, we must have $C^{II}_s = 0$ for all $s$. 

The last case in Table \ref{table} (synchronous and extreme), however, must be analyzed in more detail. 
Starting with $m\neq 0$, Eq.~\eqref{GIII} implies that $G_s^{I,II}(r)$ is regular at $r=M$ if and only if $\mathrm{Re}(-1/2 + im - s \pm \delta_s) > 0$ (as before, the upper/lower sign refers to I/II). We have two possibilities:

\begin{itemize}

\item[(i)] If $\delta_s$ is purely imaginary ($\delta_s^2 < 0$) this condition becomes $s < - 1/2$ for both $G_s^{I}$ and $G_s^{II}$, and will be automatically satisfied for $s=-1$ and $s=-2$ (and never satisfied for $s=0,+1,+2$). Therefore, to avoid irregular solutions at the horizon, we need $C^{I}_s = C^{II}_s = 0$ for $s \geq 0$. Again, due to the Taukoslky-Starobinski identities, this implies $C^{I}_s = C^{II}_s = 0$ for all $s$, so that the only regular solution is the trivial one. Thus, we can rule out the possibility of $\delta_s^2 < 0$ for synchronous frequencies with $m \neq 0$.

\item[(ii)] If, on the other hand, $\delta_s$ is real ($\delta_s^2 > 0$), the regularity condition becomes $-1/2 - s \pm \delta_s > 0$. For the minus sign (corresponding to $G^{II}_s$), it is easy to see that this condition will not be fulfilled for $s=0$, $s=1$, and $s=2$. Therefore, in order to allow $\Gamma_s$ to be regular at the horizon, we need $C^{II}_s = 0$. (As before, by virtue of the Teukolsky-Starobinski relations, this must hold for any $s$, not only for $s\geq 0$). Turning our attention to the plus sign (corresponding to $G^{I}_s$), to guarantee nontrivial regular solutions we must have both $\delta_s > 1/2 + s$ and $\delta_s > 1/2 - s$ for any given $s$ (or, equivalently, $\delta_s > 1/2 + |s|$), meaning that both $G^{I}_s$ and $G^{I}_{-s}$ are regular at the horizon. If only one of the solutions, let's say $G^{I}_s$, were regular, then $C_{s}^I$ would vanish, leading, because of \eqref{srelation}, to the trivial solution.

\end{itemize}

 In sum, we are left with the conditions $\delta_s^2 > 0$ and $\delta_s > 1/2 + |s|$ to guarantee regularity at the event horizon. With our convention $\mathrm{Re}(\delta_s) \ge 0$, and since we need to worry only about the nonnegative spin-weights, a more compact way to express this pair of inequalities is through the single condition 
\be \label{conditionx}
\delta_s ^2 > (1/2 + s)^2, \qquad s \geq 0,
\ee
which is equivalent to \eqref{reg_condition}.

Finally, when $m=0$, the synchronous frequency becomes $\omega=0$, and the parameters $\lambda_{s \ell m}(m/2)$ and $\delta_s$ reduce, respectively, to $(\ell - s)(\ell + s + 1)$ and $\ell+1/2$. In this case, we have $G^I_s(r) \sim (r-M)^{\ell - s}$ and $G^{II}_s(r) \sim (r-M)^{-\ell -s-1}$. Since $\ell \ge |s|$, the first solution is always regular at the horizon, while the second one is always irregular.    
%

%================
\section{Do zero-damped modes exist for extremal black holes?}
%================
Typically, since one does not know the quasinormal frequencies a priori, the Teukolsky equation cannot be straightforwardly integrated. Finding the QNMs becomes an eigenvalue problem, which requires a numerical method (the continued fraction method~\cite{Leaver:1985ax} is usually chosen). In Ref.~\cite{Richartz:2015saa}, an extensive search for QNMs of an extremal Kerr black hole was performed using a modification of the continued fraction method~\cite{Onozawa:1995vu}. Unfortunately, the method fails when the frequency of the modes is exactly equal to the synchronous frequency, and therefore it cannot be used to determine whether ZDMs exist for extremal black holes. Here, however, we do not need such an extensive search because we focus on (known) frequencies $\omega = m \Omega_h$, with $m \in \mathbb{Z}$. In these special cases, we can simply plug the frequencies into the radial Teukolsky equation and solve it to verify whether both boundary conditions (at the horizon and at infinity) can be simultaneously satisfied. This can be done analytically, since the Teukolsky equation for synchronous perturbations around an extremal black hole is exactly solvable in terms of confluent hypergeometric functions~\cite{1974ApJ...193..443T,2010PhLA..374.2901H}.

The simplest case is $m=0$, for which the most general solution of the Teukolsky equation is simply
\be R_s = A_s(r-M)^{\ell-s} + B_s(r-M)^{-\ell-s-1}, \ee
where $A_s$ and $B_s$ are constants. As explained before, $B_s$ must be zero to have well-behaved solutions at the horizon. If $A_s\neq 0$, however, $R_s(r)$ [and also $G_s(r)$] will diverge when $r \rightarrow \infty$ unless $\ell = s =0$ (note that $s=\ell > 0$ is ruled out, because the coexisting solution for $s = -\ell$ would be ill-behaved). In conclusion, if $\omega=m=0$, nonzero well-behaved perturbations are only possible for scalar modes with $\ell=0$, in which case the corresponding solution is simply a constant.

The analysis for nonaxisymmetric modes is more involved. In fact, if $m \neq 0$, the most general solution is 
\begin{align}
R_{s}(r) =&(r-M)^{-1 -s} \left[A_s M\left(-im + s, \delta_s; \frac{im(r-M)}{M}\right) \right. \nonumber\\
&\left. + B_s M\left(-im + s, -\delta_s;  \frac{im(r-M)}{M}\right) \right], \label{1f1}
\end{align}
where $A_s$ and $B_s$ are again constants, and $M(\beta,\gamma;z)$ is the Whittaker M function. Since $M(\beta,\gamma;z)\rightarrow z^{\gamma + 1/2}$ as $z \rightarrow 0$, in complete agreement with \eqref{near5}, we deduce that $A_s$ is associated with $R^I_s$ (and $G^I_s$), while $B_s$ is associated with $R^{II}_s$ (and $G^{II}_s$). The analysis of the previous section has already shown us that nontrivial regular solutions are only possible when $B_s=0$ and $\delta_s > 1/2 + |s|$. To check if the boundary condition of no incoming waves from $r \rightarrow \infty$ can be simultaneously satisfied, we use the asymptotic series expansion of the confluent hypergeometric function to find that, far away from the black hole, the solution when $B_s=0$ is given by 
\be
R_{s}(r) = C_s r^{-1-2s+im}e^{\frac{i m r}{2M}} + D_s r^{-1-im} e^{-\frac{i m r}{2M}}, \label{far2}
\ee
where
\begin{align}
C_s = A_s \left(\frac{im}{M}\right)^{im - s} \frac{e^{-i\frac{m}{2}}\Gamma(1+2\delta_s)}{\Gamma\left(\frac{1}{2}+im - s +\delta_s\right)}
\end{align}
and 
\begin{align}
D_s =  A_s\left(-\frac{im}{M}\right)^{-im + s} \frac{e^{i\frac{m}{2}}(-i)^{-1-2\delta_s}\Gamma(1+2\delta_s)}{\Gamma\left(\frac{1}{2}-im + s +\delta_s\right)}.
\end{align}

So we have a wave-like solution \eqref{far2} consisting of a superposition of ingoing and outgoing parts, and it is impossible to satisfy both boundary conditions at the same time. Hence ZDMs are not allowed for extremal Kerr black holes as natural oscillations (normal modes): \textit{the synchronous frequencies always correspond to scattering modes.} 

We can associate reflection and transmission coefficients with these modes.  We first note that the quantity $W_s=H_s (d H_{-s}^* /dr)  - (d H_{s}/dr) H_{-s}^*$, where $H_{s}(r) = (r-M)^{s+1}R_s(r)$, is independent of $r$. By plugging in the asymptotic expansions of \eqref{1f1} into $W_s$, if $m\neq0$ and $\delta_s^2>0$ -- as required by Eq.~\eqref{conditionx} -- we obtain the following relation:
 \be \label{wcons}
D_s D_{-s}^* - C_s C_{-s}^*  = \frac{2i\delta_s}{M}\left(e^{-i \pi \delta_s} B_s A_{-s}^*  - e^{i \pi \delta_s}A_s B_{-s}^* \right). 
  \ee
In the case of regular solutions we have $B_s = 0$ and, therefore, the right-hand side of the expression above vanishes. Using a clever insight by Teukolsky and Press~\cite{1974ApJ...193..443T}, the reflection coefficient for the scattering of a spin-$s$ perturbation is given by
\be \label{reflec}
\mathcal{R}_s = \left|\frac{C_s C_{-s}}{D_s D_{-s}}\right| = 1,
\ee
where the last equality is a consequence of $B_s$ = 0 in \eqref{wcons}. The fact that $\mathcal{R}_s=1$ means that synchronous modes are purely reflected by the extremal black hole, as one would have concluded from Eq.~\eqref{reflec0} solely from naive continuity arguments.

%================
\section{Stability of extremal black holes}
%================

We have already noticed that, unless condition \eqref{reg_condition} is met, scalar, electromagnetic, and gravitational perturbations with the synchronous frequency  will be ill-behaved at the event horizon of an extremal black hole. However, we still have to check for which sets of quantum numbers $(s,\ell, m)$ this condition is satisfied, thus allowing regular, synchronous partial waves to scatter off an extremal Kerr black hole. 

\begin{figure}[ht!]
\begin{center}
\begin{tabular}{c}
\includegraphics[width=8.0cm]{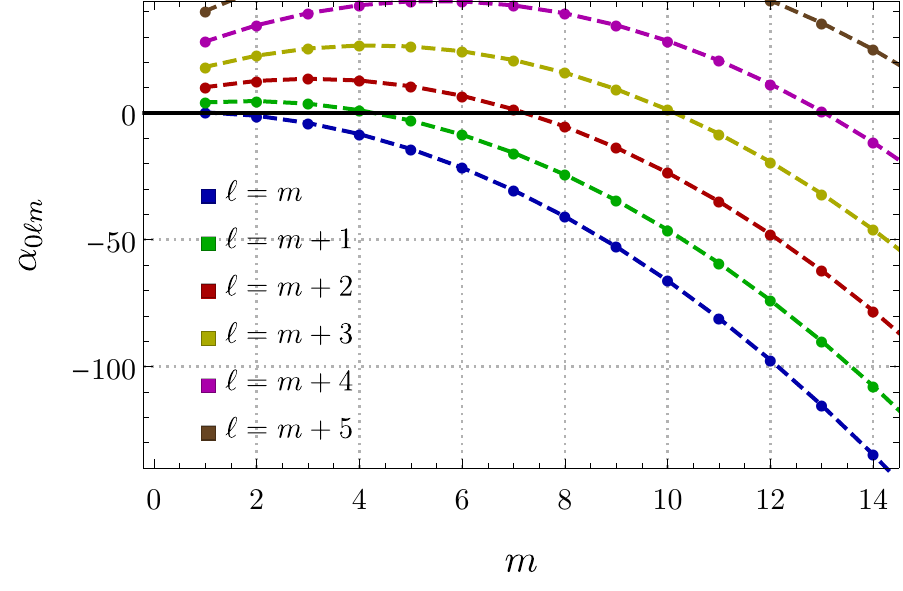}\\
\includegraphics[width=8.0cm]{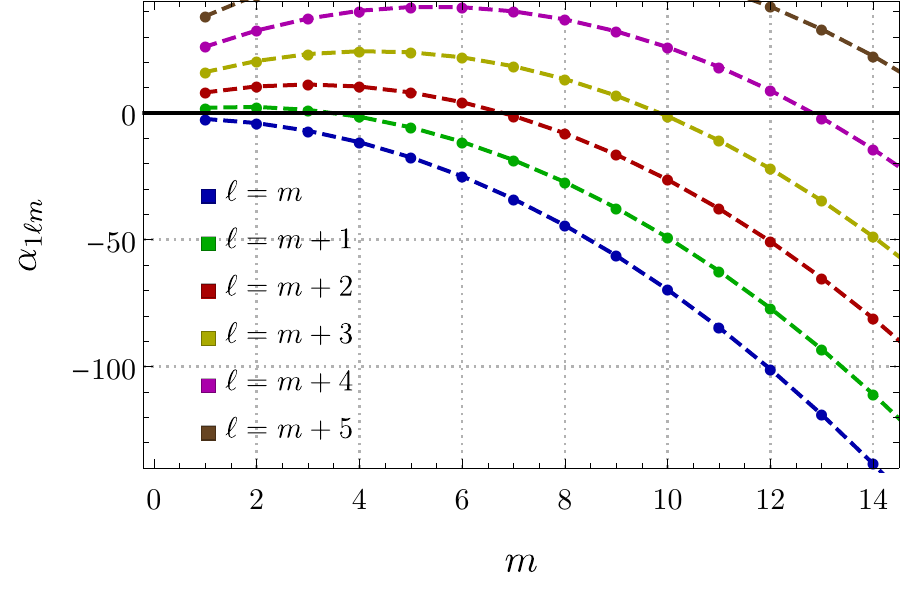}\\
\includegraphics[width=8.0cm]{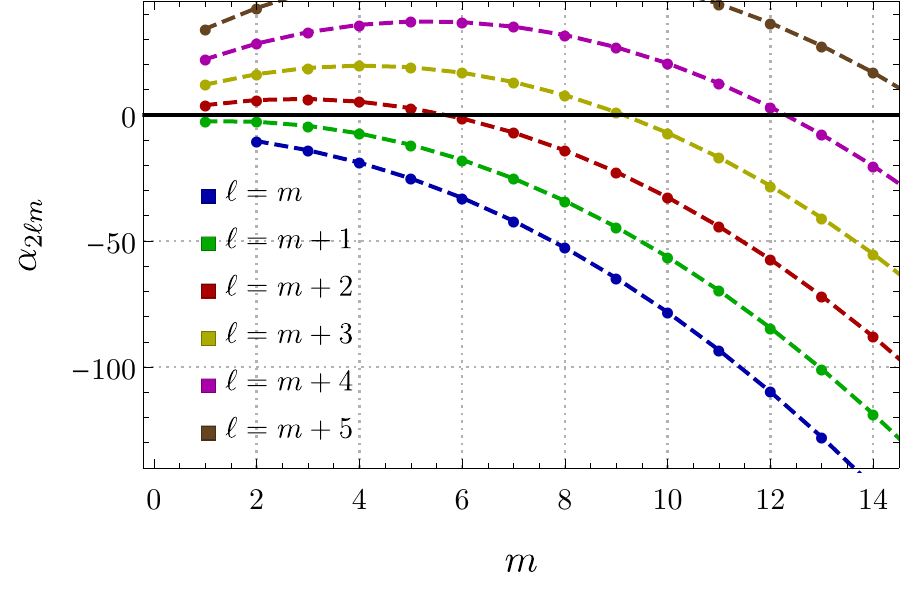}
\end{tabular}
\caption{The value of $\alpha_{0 \ell m}$ (top panel), $\alpha_{1 \ell m}$ (middle panel), and $\alpha_{2 \ell m}$ (bottom panel) as a function of $m$ [$\alpha_{s \ell m}$ is defined in \eqref{reg_condition}]. In each panel, the dashed curves correspond (from bottom to top) to $\ell = m$, $\dots$, $\ell = m + 5$. Points above the horizontal axis (solid black line) satisfy $\alpha_{s \ell m} > 0$, corresponding to sets $(\ell,m)$ for which at least one of the independent solutions of the Teukolsky equation is regular at the event horizon. Points  above the horizontal axis and below the line  $\alpha_{s \ell m}  = 2(1+s)$ (not shown), correspond to regular solutions whose first radial derivative blows-up at the event horizon.  \label{fig1}
}
\end{center}
\end{figure}

By virtue of \eqref{reg_condition}, only a discrete set of values is allowed for $\alpha_{s \ell m}$.
In general, the separation constant $\lambda_{s \ell m}(m/2)$ (and, consequently, $\alpha_{s \ell m}$) can only be determined numerically. 
The exception is $m=0$, in which case we have already seen that the solution is always ill-behaved unless $\ell=s=0$. For $m\neq 0$, the easiest way to check if regular solutions at the horizon are possible is to perform an extensive search over the discrete set of triples $(s,\ell,m)$. The result of this search is shown in Fig.~\ref{fig1}, where we plot the value of $\alpha_{s \ell m}$ as a function of $m$ for several values of $\ell$ for scalar (top panel), electromagnetic (middle panel) and gravitational (bottom panel) perturbations. 
The horizontal axis (black solid line) separates regular solutions from nonregular ones. It is clear that several sets of parameters, being above this axis, satisfy $\alpha_{s \ell m} > 0$, allowing regular solutions at the horizon.

\begin{figure}[ht!]
\begin{center}
\begin{tabular}{c}
\includegraphics[width=8.0cm]{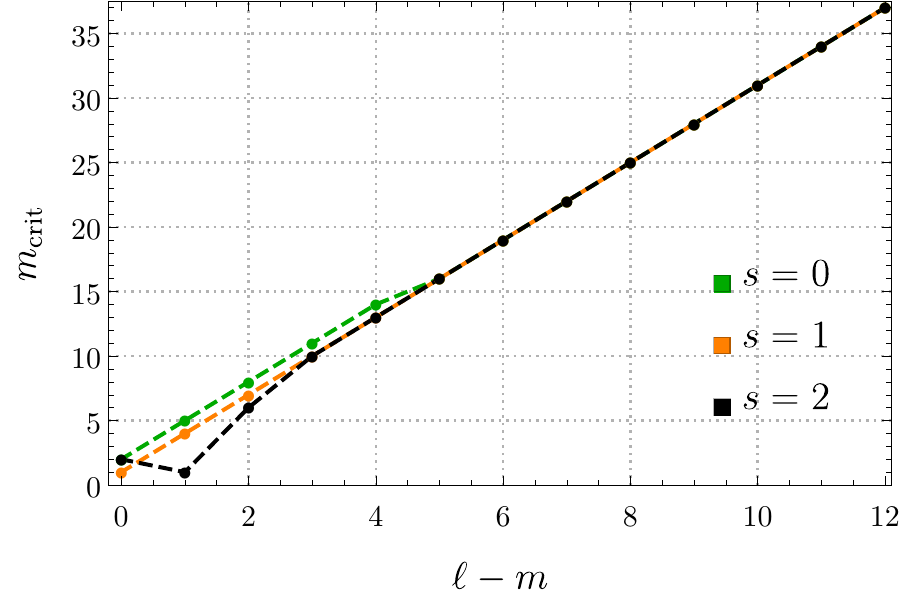}
\end{tabular}
\caption{The critical $m$ value $m_{\mathrm{crit}}$ for $s=0$ (blue line), $s=1$ (green line), and $s=2$ (red line) as a function of $\ell - m$. As $\ell - m$ increases, $m_{\rm crit}$ also increases, roughly linearly. \label{fig2}
}
\end{center}
\end{figure}

Also interesting is the fact that several other sets of $(\ell, m)$ do not satisfy this condition. From Fig.~\ref{fig1}, we can infer that, for higher values of $m$, the number of possible $\ell$ values for which the solution is irregular at the horizon increases. For instance, when $\ell = m \ge 2$ the wave function is always irregular at the horizon for scalar, electromagnetic and gravitational perturbations. The exact same thing also occurs for all gravitational perturbations when $m = \ell - 1$. Nonzero scalar modes with $m = \ell - 1$, on the other hand, will always be irregular at the horizon if $m \ge 5$. For a given type of perturbation, we define the critical $m$ value, $m_{\mathrm{crit}}$, to be the first $m$ value for which the only regular solution is the trivial one. In other words, for a given $s$, $m_{\mathrm{crit}}$ is the first integer $m$ satisfying $\alpha_{s\ell m} < 0$. We summarize our findings in Fig.~\ref{fig2}.

One could invoke the nonexistence of ZDMs to conclude that any perturbation with frequency $\omega = m \Omega_h$ will not be spontaneously excited for extremal black holes (in other words, that the only allowed solution to the wave equation is \eqref{1f1} with $A_s=B_s=0$). Nonetheless, this reasoning does not prevent a gedanken scattering experiment in which an asymptotic observer produces an external perturbation of frequency $\omega=m \Omega_h$ and sends it towards the black hole.

Far away from the black hole it is impossible to distinguish the Kerr spacetime from flat spacetime, and therefore there is no difficulty in producing perturbations with one of the synchronous frequencies. One can argue that this is a very special (possibly unphysical) situation, since it apparently requires a fine-tuning of the frequency. To avoid this type of objection, we consider an initial wavepacket that includes a continuum of frequencies around a given synchronous frequency, for instance $\psi = \sum_{\ell,m} \int_{-\infty}^{\infty}a(\omega)\Gamma_{s\omega \ell m} d\omega$, where the function $a(\omega)$ has compact support and is peaked around $\omega=m\Omega_h$ for a given $m\in \mathbb{Z}^*$.  

 If the initial wave packet includes at least one of the pathological frequencies, the corresponding wave function will necessarily diverge at the event horizon. As a consequence, only a selection of modes $(\ell,m)$, with $\omega=m\Omega_h$, are allowed in the initial wave packet. The corresponding behavior at the horizon, given by \eqref{GIII}, is $G^{I}_s \sim (r-M)^{-\frac{1}{2}+im - s + \delta_s}$, with $\delta_s > \frac{1}{2}+s$. 
The function $G^{I}_s$ is regular at the horizon, but its $n$-th radial derivative
\be  \label{are_div}
\frac{d^n G^{I}_{s}}{dr^n} \sim (r-M)^{-n-\frac{1}{2}+im - s + \delta_s}
\ee
will diverge if $n > \delta_s - s - 1/2$. Invoking the arguments of Ref.~\cite{Lucietti:2012xr}, divergences in the radial derivatives of the perturbation fields at the horizon will produce divergences of the perturbations themselves as $v \rightarrow \infty$. The worst case occurs for those sets of parameters $(\ell,m)$ that satisfy $|s| + 1/2 < \delta_s  < |s| + 3/2$, for which the first derivative already blows up at the horizon. In terms of $\alpha_{s\ell m}$, this condition is simply $ 0 < \alpha _{s \ell m} < 2(1+|s|)$. For instance, this is what happens when $(s,\ell,m)=(0,1,1)$, for which $\delta_s \approx 0.671$, or when $(s,\ell,m)=(2,3,1)$, for which $\delta_s \approx 3.175$. Several, possibly infinite, combinations exist for each spin parameter.

The result above is basically the Aretakis instability for extreme black holes, which states that, for initial data with support away from the horizon, sufficently high-order transversal derivatives blow up (at least polynomially) in time along the future event horizon. In other words, we have just shown, in a simple and intuitive way, that sufficently high-order derivatives of any wave packet that includes regular synchronous frequencies will diverge as the wave approaches the event horizon. The divergence we obtained in Eq.~\eqref{are_div} exactly matches the one obtained in \cite{Casals:2016mel} for scalar fields through an analysis of the branch points of the causal Green function. Note the importance of both the extremality and the synchronous conditions in our analysis, since any other combination (as seen in Table \ref{table}) will produce regular solutions at the horizon. Ref.~\cite{Casals:2016mel}, on the other hand, while working in the extremal case, never dealt with calculations performed exactly at the synchronous frequency, since all of their results were derived from nonsynchronous frequencies in the limit $\omega \rightarrow m\Omega_h$.   
   
%================
\section{Final Remarks}
%=========

The synchronous frequencies $\omega = m\Omega_h$ lead to a very peculiar behavior of the wave function at the event horizon of an extremal black hole. First of all, our analysis confirms the suggestion of Ref.~\cite{Casals:2016mel} that ZDMs do not constitute quasinormal type resonances of extremal black holes, being \textit{always} associated with scattering modes. Unlike damped modes of a near-extremal black hole, which are continuous at extremality~\cite{Richartz:2015saa}, ZDMs are not: ZDMs do not exist for extremal black holes as solutions of the wave equation with QNM boundary conditions. This peculiar behavior is somewhat reminiscent of the subtle nature of the boundary conditions for the algebraically special modes, which (as first discussed in~\cite{MaassenvandenBrink:2000iwh}) are not genuine QNMs, but rather correspond to total transmission/reflection modes with different boundary conditions (see also~\cite{Berti:2003jh,Cook:2014cta,Berti:2009kk}). 

Moreover, for some set of parameters $(s,\ell,m)$, the scattering modes associated with synchronous frequencies are either identically zero or diverge at the event horizon. If we rule out these modes, the remaining ones can be well-behaved at the event horizon, but higher order derivatives will necessarily diverge, as in the Aretakis instability. Even though this instability seems to require a fine tuning of the frequency, this behavior becomes generic as soon as one considers wave packets instead of single modes, providing a simple intuitive understanding of the Aretakis instability.

Furthermore, if backreaction is allowed, the simple argument below (inspired by Refs.~\cite{Glampedakis:2001js, Benone:2014ssa}) suggests that there might exist a synchronization mechanism that drives all modes to the synchronous limit. Assume indeed that a continuous flux of modes with azimuthal number $m>0$ and positive frequency $\omega \neq m \Omega_h$ incides upon an extremal Kerr black hole of mass $M$. As each mode is absorbed by the black hole, it will change its mass by $\delta M$ and its angular momentum by $\delta J$ (recall that the angular momentum $J$ of a Kerr black hole is given by $J=aM$). 

On the other hand, the average fluxes of energy $E_{\rm in}$ and angular momentum $L_{\rm in}$ far away from the black hole are given by
\be
E_{\rm in} \propto \omega (1-\mathcal{R}_{s}), \qquad L_{\rm in} \propto m (1-\mathcal{R}_{s}),
\ee    
where $\mathcal{R}_{s}$ is again the reflection coefficient of the scattering process. By conservation of energy, we must then have $\delta M \propto \omega (1-\mathcal{R}_{s})$ and $\delta J \propto m (1-\mathcal{R}_{s})$. 
If the incident wave has, initially, a frequency $\omega > m\Omega_h$, then $\mathcal{R}_{s} < 1$. As a result of the scattering process, some mass and angular momentum are transferred to the black hole. Consequently the angular velocity of the black hole will increase, approaching $\omega = m \Omega_h$. If, on the other hand, $\omega < m \Omega_h$ initially, the scattering will be superradiant ($\mathcal{R}_{s} > 1$). The net result of the process is that the reflected wave will carry away part of the mass and angular momentum of the black hole, decreasing its angular velocity until $\omega = m \Omega_h$. Note, however, that a fully nonlinear analysis of the problem is required to determine if this is indeed the case (especially for extremal black holes, since one must worry about maintaining extremality). In this respect, we remark that a recent nonlinear analysis showed synchronisation of bound perturbations can indeed be achieved dynamically~\cite{East:2017ovw}.

Finally, we remark that massless fermionic fields ($s=\pm 1/2$ and $s=\pm 3/2$) around a Kerr black hole are also described by the Teukolsky equation~\cite{1973ApJ...185..635T,Unruh:1973bda,TorresdelCastillo:1990aw}, leading (with minor differences) to the same conclusions we obtained for bosonic fields. We also point out that the behavior of charged fields $(s=0,\, \pm 1/2)$ around an extremal Reissner-Nordstr\"om black hole is completely analogous to the rotating case~\cite{Page:1976jj,Gibbons:1975kk,2011PhRvD..84j4021R,Degollado:2013eqa,Sampaio:2014swa,Richartz:2014jla}. The synchronous frequency is now $\omega = q\Phi_h$, where $q$ is the charge of the field and $\Phi_h$ is the electric potential at the event horizon (for an extremal Reissner-Nordstr\"om black hole, $\Phi_h=1$). Instead of $\delta_s = \sqrt{(1/2 + s)^2 - 7m^2/4 + \lambda_{s \ell m}(m/2)}$, we now have $\delta_s = \sqrt{(1/2 + s)^2 - q^2M^2 + \lambda_{s \ell m}(0)}$, which can be further simplified, due to the spherical symmetry, to $\delta_s = \sqrt{(1/2 + \ell)^2 - q^2M^2}$ . When $qM < \ell+1/2$, the trivially vanishing solution is the only regular solution for the synchronous frequency. We therefore require $qM > \ell+1/2$, so that at least one nonzero solution which is regular at the horizon exists, corresponding to $R_s \sim (r-M)^{-1/2 - s 
+ \delta_s}$. As in the extremal Kerr case, sufficiently high-order radial derivatives will always diverge like $\sim (r-M)^{-n -1/2 - s + \delta_s}$, where $n$ is the number of derivatives, as in the original derivation by Aretakis.

\bigskip

%================
\acknowledgments
%================

The authors would like to thank A.~Zimmerman, E.~Poisson, K.~Chatziioannou, and P.~Zimmerman for illuminating discussions regarding Eq.~\eqref{teukorad2}.
M.R.~is grateful to A.~Nerozzi, A.~Saa, A.~Zimmerman, and J.~P.~Pitelli for enlightning discussions. M.R.~ackowledges support from the S\~ao Paulo Research Foundation (FAPESP), Grant No.~2013/09357-9, and from the Fulbright Visiting Scholars Program. M.R.~is also grateful to the University of Mississippi for hospitality while part of this research was being conducted. C.A.R.H.~acknowledges  funding from  the  FCT-IF  programme. E.B.~is supported by NSF Grants No.~PHY-1607130 and AST-1716715, and by FCT contract IF/00797/2014/CP1214/CT0012 under the IF2014 Programme. This  work  was  partially supported by the H2020-MSCA-RISE-2015 Grant No.~StronGrHEP-690904, and by the CIDMA  project UID/MAT/04106/2013.

\bibliography{zdm}

\begin{thebibliography}{87}%
\makeatletter
\providecommand \@ifxundefined [1]{%
 \@ifx{#1\undefined}
}%
\providecommand \@ifnum [1]{%
 \ifnum #1\expandafter \@firstoftwo
 \else \expandafter \@secondoftwo
 \fi
}%
\providecommand \@ifx [1]{%
 \ifx #1\expandafter \@firstoftwo
 \else \expandafter \@secondoftwo
 \fi
}%
\providecommand \natexlab [1]{#1}%
\providecommand \enquote  [1]{``#1''}%
\providecommand \bibnamefont  [1]{#1}%
\providecommand \bibfnamefont [1]{#1}%
\providecommand \citenamefont [1]{#1}%
\providecommand \href@noop [0]{\@secondoftwo}%
\providecommand \href [0]{\begingroup \@sanitize@url \@href}%
\providecommand \@href[1]{\@@startlink{#1}\@@href}%
\providecommand \@@href[1]{\endgroup#1\@@endlink}%
\providecommand \@sanitize@url [0]{\catcode `\\12\catcode `\$12\catcode
  `\&12\catcode `\#12\catcode `\^12\catcode `\_12\catcode `\%12\relax}%
\providecommand \@@startlink[1]{}%
\providecommand \@@endlink[0]{}%
\providecommand \url  [0]{\begingroup\@sanitize@url \@url }%
\providecommand \@url [1]{\endgroup\@href {#1}{\urlprefix }}%
\providecommand \urlprefix  [0]{URL }%
\providecommand \Eprint [0]{\href }%
\providecommand \doibase [0]{http://dx.doi.org/}%
\providecommand \selectlanguage [0]{\@gobble}%
\providecommand \bibinfo  [0]{\@secondoftwo}%
\providecommand \bibfield  [0]{\@secondoftwo}%
\providecommand \translation [1]{[#1]}%
\providecommand \BibitemOpen [0]{}%
\providecommand \bibitemStop [0]{}%
\providecommand \bibitemNoStop [0]{.\EOS\space}%
\providecommand \EOS [0]{\spacefactor3000\relax}%
\providecommand \BibitemShut  [1]{\csname bibitem#1\endcsname}%
\let\auto@bib@innerbib\@empty
%</preamble>
\bibitem [{\citenamefont {Kokkotas}\ and\ \citenamefont
  {Schmidt}(1999)}]{Kokkotas:1999bd}%
  \BibitemOpen
  \bibfield  {author} {\bibinfo {author} {\bibfnamefont {K.~D.}\ \bibnamefont
  {Kokkotas}}\ and\ \bibinfo {author} {\bibfnamefont {B.~G.}\ \bibnamefont
  {Schmidt}},\ }\href {\doibase 10.12942/lrr-1999-2} {\bibfield  {journal}
  {\bibinfo  {journal} {Living Rev. Rel.}\ }\textbf {\bibinfo {volume} {2}},\
  \bibinfo {pages} {2} (\bibinfo {year} {1999})},\ \Eprint
  {http://arxiv.org/abs/gr-qc/9909058} {arXiv:gr-qc/9909058 [gr-qc]}
  \BibitemShut {NoStop}%
%%CITATION = GR-QC/9909058;%%
\bibitem [{\citenamefont {Nollert}(1999)}]{Nollert:1999ji}%
  \BibitemOpen
  \bibfield  {author} {\bibinfo {author} {\bibfnamefont {H.-P.}\ \bibnamefont
  {Nollert}},\ }\href {\doibase 10.1088/0264-9381/16/12/201} {\bibfield
  {journal} {\bibinfo  {journal} {Class. Quant. Grav.}\ }\textbf {\bibinfo
  {volume} {16}},\ \bibinfo {pages} {R159} (\bibinfo {year}
  {1999})}\BibitemShut {NoStop}%
%%CITATION = CQGRD,16,R159;%%
\bibitem [{\citenamefont {Berti}\ \emph {et~al.}(2009)\citenamefont {Berti},
  \citenamefont {Cardoso},\ and\ \citenamefont {Starinets}}]{Berti:2009kk}%
  \BibitemOpen
  \bibfield  {author} {\bibinfo {author} {\bibfnamefont {E.}~\bibnamefont
  {Berti}}, \bibinfo {author} {\bibfnamefont {V.}~\bibnamefont {Cardoso}}, \
  and\ \bibinfo {author} {\bibfnamefont {A.~O.}\ \bibnamefont {Starinets}},\
  }\href {\doibase 10.1088/0264-9381/26/16/163001} {\bibfield  {journal}
  {\bibinfo  {journal} {Class. Quant. Grav.}\ }\textbf {\bibinfo {volume}
  {26}},\ \bibinfo {pages} {163001} (\bibinfo {year} {2009})},\ \Eprint
  {http://arxiv.org/abs/0905.2975} {arXiv:0905.2975 [gr-qc]} \BibitemShut
  {NoStop}%
%%CITATION = ARXIV:0905.2975;%%
\bibitem [{\citenamefont {Konoplya}\ and\ \citenamefont
  {Zhidenko}(2011)}]{Konoplya:2011qq}%
  \BibitemOpen
  \bibfield  {author} {\bibinfo {author} {\bibfnamefont {R.~A.}\ \bibnamefont
  {Konoplya}}\ and\ \bibinfo {author} {\bibfnamefont {A.}~\bibnamefont
  {Zhidenko}},\ }\href {\doibase 10.1103/RevModPhys.83.793} {\bibfield
  {journal} {\bibinfo  {journal} {Rev. Mod. Phys.}\ }\textbf {\bibinfo {volume}
  {83}},\ \bibinfo {pages} {793} (\bibinfo {year} {2011})},\ \Eprint
  {http://arxiv.org/abs/1102.4014} {arXiv:1102.4014 [gr-qc]} \BibitemShut
  {NoStop}%
%%CITATION = ARXIV:1102.4014;%%
\bibitem [{\citenamefont {{Thorne}}(1974)}]{1974ApJ...191..507T}%
  \BibitemOpen
  \bibfield  {author} {\bibinfo {author} {\bibfnamefont {K.~S.}\ \bibnamefont
  {{Thorne}}},\ }\href {\doibase 10.1086/152991} {\bibfield  {journal}
  {\bibinfo  {journal} {\apj}\ }\textbf {\bibinfo {volume} {191}},\ \bibinfo
  {pages} {507} (\bibinfo {year} {1974})}\BibitemShut {NoStop}%
\bibitem [{\citenamefont {Sadowski}\ \emph {et~al.}(2011)\citenamefont
  {Sadowski}, \citenamefont {Bursa}, \citenamefont {Abramowicz}, \citenamefont
  {Kluzniak}, \citenamefont {Lasota}, \citenamefont {Moderski},\ and\
  \citenamefont {Safarzadeh}}]{Sadowski:2011ka}%
  \BibitemOpen
  \bibfield  {author} {\bibinfo {author} {\bibfnamefont {A.}~\bibnamefont
  {Sadowski}}, \bibinfo {author} {\bibfnamefont {M.}~\bibnamefont {Bursa}},
  \bibinfo {author} {\bibfnamefont {M.}~\bibnamefont {Abramowicz}}, \bibinfo
  {author} {\bibfnamefont {W.}~\bibnamefont {Kluzniak}}, \bibinfo {author}
  {\bibfnamefont {J.-P.}\ \bibnamefont {Lasota}}, \bibinfo {author}
  {\bibfnamefont {R.}~\bibnamefont {Moderski}}, \ and\ \bibinfo {author}
  {\bibfnamefont {M.}~\bibnamefont {Safarzadeh}},\ }\href {\doibase
  10.1051/0004-6361/201116702} {\bibfield  {journal} {\bibinfo  {journal}
  {Astron. Astrophys.}\ }\textbf {\bibinfo {volume} {532}},\ \bibinfo {pages}
  {A41} (\bibinfo {year} {2011})},\ \Eprint {http://arxiv.org/abs/1102.2456}
  {arXiv:1102.2456 [astro-ph.HE]} \BibitemShut {NoStop}%
%%CITATION = ARXIV:1102.2456;%%
\bibitem [{\citenamefont {Strominger}\ and\ \citenamefont
  {Vafa}(1996)}]{Strominger:1996sh}%
  \BibitemOpen
  \bibfield  {author} {\bibinfo {author} {\bibfnamefont {A.}~\bibnamefont
  {Strominger}}\ and\ \bibinfo {author} {\bibfnamefont {C.}~\bibnamefont
  {Vafa}},\ }\href {\doibase 10.1016/0370-2693(96)00345-0} {\bibfield
  {journal} {\bibinfo  {journal} {Phys. Lett.}\ }\textbf {\bibinfo {volume}
  {B379}},\ \bibinfo {pages} {99} (\bibinfo {year} {1996})},\ \Eprint
  {http://arxiv.org/abs/hep-th/9601029} {arXiv:hep-th/9601029 [hep-th]}
  \BibitemShut {NoStop}%
%%CITATION = HEP-TH/9601029;%%
\bibitem [{\citenamefont {Guica}\ \emph {et~al.}(2009)\citenamefont {Guica},
  \citenamefont {Hartman}, \citenamefont {Song},\ and\ \citenamefont
  {Strominger}}]{Guica:2008mu}%
  \BibitemOpen
  \bibfield  {author} {\bibinfo {author} {\bibfnamefont {M.}~\bibnamefont
  {Guica}}, \bibinfo {author} {\bibfnamefont {T.}~\bibnamefont {Hartman}},
  \bibinfo {author} {\bibfnamefont {W.}~\bibnamefont {Song}}, \ and\ \bibinfo
  {author} {\bibfnamefont {A.}~\bibnamefont {Strominger}},\ }\href {\doibase
  10.1103/PhysRevD.80.124008} {\bibfield  {journal} {\bibinfo  {journal} {Phys.
  Rev.}\ }\textbf {\bibinfo {volume} {D80}},\ \bibinfo {pages} {124008}
  (\bibinfo {year} {2009})},\ \Eprint {http://arxiv.org/abs/0809.4266}
  {arXiv:0809.4266 [hep-th]} \BibitemShut {NoStop}%
%%CITATION = ARXIV:0809.4266;%%
\bibitem [{\citenamefont {Wiltshire}\ \emph {et~al.}(2009)\citenamefont
  {Wiltshire}, \citenamefont {Visser},\ and\ \citenamefont
  {Scott}}]{Wiltshire:2009zza}%
  \BibitemOpen
  \bibfield  {author} {\bibinfo {author} {\bibfnamefont {D.~L.}\ \bibnamefont
  {Wiltshire}}, \bibinfo {author} {\bibfnamefont {M.}~\bibnamefont {Visser}}, \
  and\ \bibinfo {author} {\bibfnamefont {S.~M.}\ \bibnamefont {Scott}},\ }\href
  {http://www.cambridge.org/catalogue/catalogue.asp?isbn=9780521885126} {\emph
  {\bibinfo {title} {{The Kerr spacetime: Rotating black holes in general
  relativity}}}}\ (\bibinfo  {publisher} {Cambridge University Press},\
  \bibinfo {year} {2009})\BibitemShut {NoStop}%
%%CITATION = INSPIRE-831262;%%
\bibitem [{\citenamefont {Whiting}(1989)}]{Whiting:1988vc}%
  \BibitemOpen
  \bibfield  {author} {\bibinfo {author} {\bibfnamefont {B.~F.}\ \bibnamefont
  {Whiting}},\ }\href {\doibase 10.1063/1.528308} {\bibfield  {journal}
  {\bibinfo  {journal} {J. Math. Phys.}\ }\textbf {\bibinfo {volume} {30}},\
  \bibinfo {pages} {1301} (\bibinfo {year} {1989})}\BibitemShut {NoStop}%
%%CITATION = JMAPA,30,1301;%%
\bibitem [{\citenamefont {{Press}}\ and\ \citenamefont
  {{Teukolsky}}(1973)}]{1973ApJ...185..649P}%
  \BibitemOpen
  \bibfield  {author} {\bibinfo {author} {\bibfnamefont {W.~H.}\ \bibnamefont
  {{Press}}}\ and\ \bibinfo {author} {\bibfnamefont {S.~A.}\ \bibnamefont
  {{Teukolsky}}},\ }\href {\doibase 10.1086/152445} {\bibfield  {journal}
  {\bibinfo  {journal} {\apj}\ }\textbf {\bibinfo {volume} {185}},\ \bibinfo
  {pages} {649} (\bibinfo {year} {1973})}\BibitemShut {NoStop}%
\bibitem [{\citenamefont {{Detweiler}}\ and\ \citenamefont
  {{Ipser}}(1973)}]{1973ApJ...185..675D}%
  \BibitemOpen
  \bibfield  {author} {\bibinfo {author} {\bibfnamefont {S.~L.}\ \bibnamefont
  {{Detweiler}}}\ and\ \bibinfo {author} {\bibfnamefont {J.~R.}\ \bibnamefont
  {{Ipser}}},\ }\href {\doibase 10.1086/152446} {\bibfield  {journal} {\bibinfo
   {journal} {\apj}\ }\textbf {\bibinfo {volume} {185}},\ \bibinfo {pages}
  {675} (\bibinfo {year} {1973})}\BibitemShut {NoStop}%
\bibitem [{\citenamefont {{Friedman}}\ and\ \citenamefont
  {{Schutz}}(1974)}]{1974PhRvL..32..243F}%
  \BibitemOpen
  \bibfield  {author} {\bibinfo {author} {\bibfnamefont {J.~L.}\ \bibnamefont
  {{Friedman}}}\ and\ \bibinfo {author} {\bibfnamefont {B.~F.}\ \bibnamefont
  {{Schutz}}},\ }\href {\doibase 10.1103/PhysRevLett.32.243} {\bibfield
  {journal} {\bibinfo  {journal} {Physical Review Letters}\ }\textbf {\bibinfo
  {volume} {32}},\ \bibinfo {pages} {243} (\bibinfo {year} {1974})}\BibitemShut
  {NoStop}%
\bibitem [{\citenamefont {Richartz}(2016)}]{Richartz:2015saa}%
  \BibitemOpen
  \bibfield  {author} {\bibinfo {author} {\bibfnamefont {M.}~\bibnamefont
  {Richartz}},\ }\href {\doibase 10.1103/PhysRevD.93.064062} {\bibfield
  {journal} {\bibinfo  {journal} {Phys. Rev.}\ }\textbf {\bibinfo {volume}
  {D93}},\ \bibinfo {pages} {064062} (\bibinfo {year} {2016})},\ \Eprint
  {http://arxiv.org/abs/1509.04260} {arXiv:1509.04260 [gr-qc]} \BibitemShut
  {NoStop}%
%%CITATION = ARXIV:1509.04260;%%
\bibitem [{\citenamefont {Dafermos}\ and\ \citenamefont
  {Rodnianski}(2010)}]{Dafermos:2010hd}%
  \BibitemOpen
  \bibfield  {author} {\bibinfo {author} {\bibfnamefont {M.}~\bibnamefont
  {Dafermos}}\ and\ \bibinfo {author} {\bibfnamefont {I.}~\bibnamefont
  {Rodnianski}},\ }in\ \href {\doibase 10.1142/9789814374552_0008} {\emph
  {\bibinfo {booktitle} {{On recent developments in theoretical and
  experimental general relativity, astrophysics and relativistic field
  theories. Proceedings, 12th Marcel Grossmann Meeting on General Relativity,
  Paris, France, July 12-18, 2009. Vol. 1-3}}}}\ (\bibinfo {year} {2010})\ pp.\
  \bibinfo {pages} {132--189},\ \bibinfo {note} {[,421(2010)]},\ \Eprint
  {http://arxiv.org/abs/1010.5137} {arXiv:1010.5137 [gr-qc]} \BibitemShut
  {NoStop}%
%%CITATION = ARXIV:1010.5137;%%
\bibitem [{\citenamefont {Dafermos}\ \emph {et~al.}(2014)\citenamefont
  {Dafermos}, \citenamefont {Rodnianski},\ and\ \citenamefont
  {Shlapentokh-Rothman}}]{Dafermos:2014cua}%
  \BibitemOpen
  \bibfield  {author} {\bibinfo {author} {\bibfnamefont {M.}~\bibnamefont
  {Dafermos}}, \bibinfo {author} {\bibfnamefont {I.}~\bibnamefont
  {Rodnianski}}, \ and\ \bibinfo {author} {\bibfnamefont {Y.}~\bibnamefont
  {Shlapentokh-Rothman}},\ }\href@noop {} {\  (\bibinfo {year} {2014})},\
  \Eprint {http://arxiv.org/abs/1402.7034} {arXiv:1402.7034 [gr-qc]}
  \BibitemShut {NoStop}%
%%CITATION = ARXIV:1402.7034;%%
\bibitem [{\citenamefont
  {Shlapentokh-Rothman}(2015)}]{Shlapentokh-Rothman:2013hza}%
  \BibitemOpen
  \bibfield  {author} {\bibinfo {author} {\bibfnamefont {Y.}~\bibnamefont
  {Shlapentokh-Rothman}},\ }\href {\doibase 10.1007/s00023-014-0315-7}
  {\bibfield  {journal} {\bibinfo  {journal} {Annales Henri Poincare}\ }\textbf
  {\bibinfo {volume} {16}},\ \bibinfo {pages} {289} (\bibinfo {year} {2015})},\
  \Eprint {http://arxiv.org/abs/1302.6902} {arXiv:1302.6902 [gr-qc]}
  \BibitemShut {NoStop}%
%%CITATION = ARXIV:1302.6902;%%
\bibitem [{\citenamefont {Aretakis}(2012)}]{Aretakis:2011gz}%
  \BibitemOpen
  \bibfield  {author} {\bibinfo {author} {\bibfnamefont {S.}~\bibnamefont
  {Aretakis}},\ }\href {\doibase 10.1016/j.jfa.2012.08.015} {\bibfield
  {journal} {\bibinfo  {journal} {J. Funct. Anal.}\ }\textbf {\bibinfo {volume}
  {263}},\ \bibinfo {pages} {2770} (\bibinfo {year} {2012})},\ \Eprint
  {http://arxiv.org/abs/1110.2006} {arXiv:1110.2006 [gr-qc]} \BibitemShut
  {NoStop}%
%%CITATION = ARXIV:1110.2006;%%
\bibitem [{\citenamefont {Aretakis}(2011{\natexlab{a}})}]{Aretakis:2011ha}%
  \BibitemOpen
  \bibfield  {author} {\bibinfo {author} {\bibfnamefont {S.}~\bibnamefont
  {Aretakis}},\ }\href {\doibase 10.1007/s00220-011-1254-5} {\bibfield
  {journal} {\bibinfo  {journal} {Commun. Math. Phys.}\ }\textbf {\bibinfo
  {volume} {307}},\ \bibinfo {pages} {17} (\bibinfo {year}
  {2011}{\natexlab{a}})},\ \Eprint {http://arxiv.org/abs/1110.2007}
  {arXiv:1110.2007 [gr-qc]} \BibitemShut {NoStop}%
%%CITATION = ARXIV:1110.2007;%%
\bibitem [{\citenamefont {Aretakis}(2011{\natexlab{b}})}]{Aretakis:2011hc}%
  \BibitemOpen
  \bibfield  {author} {\bibinfo {author} {\bibfnamefont {S.}~\bibnamefont
  {Aretakis}},\ }\href {\doibase 10.1007/s00023-011-0110-7} {\bibfield
  {journal} {\bibinfo  {journal} {Annales Henri Poincare}\ }\textbf {\bibinfo
  {volume} {12}},\ \bibinfo {pages} {1491} (\bibinfo {year}
  {2011}{\natexlab{b}})},\ \Eprint {http://arxiv.org/abs/1110.2009}
  {arXiv:1110.2009 [gr-qc]} \BibitemShut {NoStop}%
%%CITATION = ARXIV:1110.2009;%%
\bibitem [{\citenamefont {Aretakis}(2013)}]{Aretakis:2012bm}%
  \BibitemOpen
  \bibfield  {author} {\bibinfo {author} {\bibfnamefont {S.}~\bibnamefont
  {Aretakis}},\ }\href {\doibase 10.1088/0264-9381/30/9/095010} {\bibfield
  {journal} {\bibinfo  {journal} {Class. Quant. Grav.}\ }\textbf {\bibinfo
  {volume} {30}},\ \bibinfo {pages} {095010} (\bibinfo {year} {2013})},\
  \Eprint {http://arxiv.org/abs/1212.1103} {arXiv:1212.1103 [gr-qc]}
  \BibitemShut {NoStop}%
%%CITATION = ARXIV:1212.1103;%%
\bibitem [{\citenamefont {Lucietti}\ and\ \citenamefont
  {Reall}(2012)}]{Lucietti:2012sf}%
  \BibitemOpen
  \bibfield  {author} {\bibinfo {author} {\bibfnamefont {J.}~\bibnamefont
  {Lucietti}}\ and\ \bibinfo {author} {\bibfnamefont {H.~S.}\ \bibnamefont
  {Reall}},\ }\href {\doibase 10.1103/PhysRevD.86.104030} {\bibfield  {journal}
  {\bibinfo  {journal} {Phys. Rev.}\ }\textbf {\bibinfo {volume} {D86}},\
  \bibinfo {pages} {104030} (\bibinfo {year} {2012})},\ \Eprint
  {http://arxiv.org/abs/1208.1437} {arXiv:1208.1437 [gr-qc]} \BibitemShut
  {NoStop}%
%%CITATION = ARXIV:1208.1437;%%
\bibitem [{\citenamefont {Lucietti}\ \emph {et~al.}(2013)\citenamefont
  {Lucietti}, \citenamefont {Murata}, \citenamefont {Reall},\ and\
  \citenamefont {Tanahashi}}]{Lucietti:2012xr}%
  \BibitemOpen
  \bibfield  {author} {\bibinfo {author} {\bibfnamefont {J.}~\bibnamefont
  {Lucietti}}, \bibinfo {author} {\bibfnamefont {K.}~\bibnamefont {Murata}},
  \bibinfo {author} {\bibfnamefont {H.~S.}\ \bibnamefont {Reall}}, \ and\
  \bibinfo {author} {\bibfnamefont {N.}~\bibnamefont {Tanahashi}},\ }\href
  {\doibase 10.1007/JHEP03(2013)035} {\bibfield  {journal} {\bibinfo  {journal}
  {JHEP}\ }\textbf {\bibinfo {volume} {03}},\ \bibinfo {pages} {035} (\bibinfo
  {year} {2013})},\ \Eprint {http://arxiv.org/abs/1212.2557} {arXiv:1212.2557
  [gr-qc]} \BibitemShut {NoStop}%
%%CITATION = ARXIV:1212.2557;%%
\bibitem [{\citenamefont {Murata}\ \emph {et~al.}(2013)\citenamefont {Murata},
  \citenamefont {Reall},\ and\ \citenamefont {Tanahashi}}]{Murata:2013daa}%
  \BibitemOpen
  \bibfield  {author} {\bibinfo {author} {\bibfnamefont {K.}~\bibnamefont
  {Murata}}, \bibinfo {author} {\bibfnamefont {H.~S.}\ \bibnamefont {Reall}}, \
  and\ \bibinfo {author} {\bibfnamefont {N.}~\bibnamefont {Tanahashi}},\ }\href
  {\doibase 10.1088/0264-9381/30/23/235007} {\bibfield  {journal} {\bibinfo
  {journal} {Class. Quant. Grav.}\ }\textbf {\bibinfo {volume} {30}},\ \bibinfo
  {pages} {235007} (\bibinfo {year} {2013})},\ \Eprint
  {http://arxiv.org/abs/1307.6800} {arXiv:1307.6800 [gr-qc]} \BibitemShut
  {NoStop}%
%%CITATION = ARXIV:1307.6800;%%
\bibitem [{\citenamefont {Casals}\ \emph {et~al.}(2016)\citenamefont {Casals},
  \citenamefont {Gralla},\ and\ \citenamefont {Zimmerman}}]{Casals:2016mel}%
  \BibitemOpen
  \bibfield  {author} {\bibinfo {author} {\bibfnamefont {M.}~\bibnamefont
  {Casals}}, \bibinfo {author} {\bibfnamefont {S.~E.}\ \bibnamefont {Gralla}},
  \ and\ \bibinfo {author} {\bibfnamefont {P.}~\bibnamefont {Zimmerman}},\
  }\href {\doibase 10.1103/PhysRevD.94.064003} {\bibfield  {journal} {\bibinfo
  {journal} {Phys. Rev.}\ }\textbf {\bibinfo {volume} {D94}},\ \bibinfo {pages}
  {064003} (\bibinfo {year} {2016})},\ \Eprint
  {http://arxiv.org/abs/1606.08505} {arXiv:1606.08505 [gr-qc]} \BibitemShut
  {NoStop}%
%%CITATION = ARXIV:1606.08505;%%
\bibitem [{\citenamefont {{Hartle}}\ and\ \citenamefont
  {{Wilkins}}(1974)}]{1974CMaPh..38...47H}%
  \BibitemOpen
  \bibfield  {author} {\bibinfo {author} {\bibfnamefont {J.~B.}\ \bibnamefont
  {{Hartle}}}\ and\ \bibinfo {author} {\bibfnamefont {D.~C.}\ \bibnamefont
  {{Wilkins}}},\ }\href {\doibase 10.1007/BF01651548} {\bibfield  {journal}
  {\bibinfo  {journal} {Communications in Mathematical Physics}\ }\textbf
  {\bibinfo {volume} {38}},\ \bibinfo {pages} {47} (\bibinfo {year}
  {1974})}\BibitemShut {NoStop}%
\bibitem [{\citenamefont {Glampedakis}\ and\ \citenamefont
  {Andersson}(2001)}]{Glampedakis:2001js}%
  \BibitemOpen
  \bibfield  {author} {\bibinfo {author} {\bibfnamefont {K.}~\bibnamefont
  {Glampedakis}}\ and\ \bibinfo {author} {\bibfnamefont {N.}~\bibnamefont
  {Andersson}},\ }\href {\doibase 10.1103/PhysRevD.64.104021} {\bibfield
  {journal} {\bibinfo  {journal} {Phys. Rev.}\ }\textbf {\bibinfo {volume}
  {D64}},\ \bibinfo {pages} {104021} (\bibinfo {year} {2001})},\ \Eprint
  {http://arxiv.org/abs/gr-qc/0103054} {arXiv:gr-qc/0103054 [gr-qc]}
  \BibitemShut {NoStop}%
%%CITATION = GR-QC/0103054;%%
\bibitem [{\citenamefont {{Detweiler}}(1980)}]{1980ApJ...239..292D}%
  \BibitemOpen
  \bibfield  {author} {\bibinfo {author} {\bibfnamefont {S.}~\bibnamefont
  {{Detweiler}}},\ }\href {\doibase 10.1086/158109} {\bibfield  {journal}
  {\bibinfo  {journal} {\apj}\ }\textbf {\bibinfo {volume} {239}},\ \bibinfo
  {pages} {292} (\bibinfo {year} {1980})}\BibitemShut {NoStop}%
\bibitem [{\citenamefont {{Teukolsky}}\ and\ \citenamefont
  {{Press}}(1974)}]{1974ApJ...193..443T}%
  \BibitemOpen
  \bibfield  {author} {\bibinfo {author} {\bibfnamefont {S.~A.}\ \bibnamefont
  {{Teukolsky}}}\ and\ \bibinfo {author} {\bibfnamefont {W.~H.}\ \bibnamefont
  {{Press}}},\ }\href {\doibase 10.1086/153180} {\bibfield  {journal} {\bibinfo
   {journal} {\apj}\ }\textbf {\bibinfo {volume} {193}},\ \bibinfo {pages}
  {443} (\bibinfo {year} {1974})}\BibitemShut {NoStop}%
\bibitem [{\citenamefont {Detweiler}\ and\ \citenamefont
  {Ove}(1983)}]{Detweiler:1983zz}%
  \BibitemOpen
  \bibfield  {author} {\bibinfo {author} {\bibfnamefont {S.~L.}\ \bibnamefont
  {Detweiler}}\ and\ \bibinfo {author} {\bibfnamefont {R.}~\bibnamefont
  {Ove}},\ }\href {\doibase 10.1103/PhysRevLett.51.67} {\bibfield  {journal}
  {\bibinfo  {journal} {Phys. Rev. Lett.}\ }\textbf {\bibinfo {volume} {51}},\
  \bibinfo {pages} {67} (\bibinfo {year} {1983})}\BibitemShut {NoStop}%
%%CITATION = PRLTA,51,67;%%
\bibitem [{\citenamefont {Ferrari}\ and\ \citenamefont
  {Mashhoon}(1984)}]{Ferrari:1984zz}%
  \BibitemOpen
  \bibfield  {author} {\bibinfo {author} {\bibfnamefont {V.}~\bibnamefont
  {Ferrari}}\ and\ \bibinfo {author} {\bibfnamefont {B.}~\bibnamefont
  {Mashhoon}},\ }\href {\doibase 10.1103/PhysRevD.30.295} {\bibfield  {journal}
  {\bibinfo  {journal} {Phys. Rev.}\ }\textbf {\bibinfo {volume} {D30}},\
  \bibinfo {pages} {295} (\bibinfo {year} {1984})}\BibitemShut {NoStop}%
%%CITATION = PHRVA,D30,295;%%
\bibitem [{\citenamefont {Sasaki}\ and\ \citenamefont
  {Nakamura}(1990)}]{Sasaki:1989ca}%
  \BibitemOpen
  \bibfield  {author} {\bibinfo {author} {\bibfnamefont {M.}~\bibnamefont
  {Sasaki}}\ and\ \bibinfo {author} {\bibfnamefont {T.}~\bibnamefont
  {Nakamura}},\ }\href {\doibase 10.1007/BF00756835} {\bibfield  {journal}
  {\bibinfo  {journal} {Gen. Rel. Grav.}\ }\textbf {\bibinfo {volume} {22}},\
  \bibinfo {pages} {1351} (\bibinfo {year} {1990})}\BibitemShut {NoStop}%
%%CITATION = GRGVA,22,1351;%%
\bibitem [{\citenamefont {Andersson}\ and\ \citenamefont
  {Glampedakis}(2000)}]{Andersson:1999wj}%
  \BibitemOpen
  \bibfield  {author} {\bibinfo {author} {\bibfnamefont {N.}~\bibnamefont
  {Andersson}}\ and\ \bibinfo {author} {\bibfnamefont {K.}~\bibnamefont
  {Glampedakis}},\ }\href {\doibase 10.1103/PhysRevLett.84.4537} {\bibfield
  {journal} {\bibinfo  {journal} {Phys. Rev. Lett.}\ }\textbf {\bibinfo
  {volume} {84}},\ \bibinfo {pages} {4537} (\bibinfo {year} {2000})},\ \Eprint
  {http://arxiv.org/abs/gr-qc/9909050} {arXiv:gr-qc/9909050 [gr-qc]}
  \BibitemShut {NoStop}%
%%CITATION = GR-QC/9909050;%%
\bibitem [{\citenamefont {Onozawa}(1997)}]{Onozawa:1996ux}%
  \BibitemOpen
  \bibfield  {author} {\bibinfo {author} {\bibfnamefont {H.}~\bibnamefont
  {Onozawa}},\ }\href {\doibase 10.1103/PhysRevD.55.3593} {\bibfield  {journal}
  {\bibinfo  {journal} {Phys. Rev.}\ }\textbf {\bibinfo {volume} {D55}},\
  \bibinfo {pages} {3593} (\bibinfo {year} {1997})},\ \Eprint
  {http://arxiv.org/abs/gr-qc/9610048} {arXiv:gr-qc/9610048 [gr-qc]}
  \BibitemShut {NoStop}%
%%CITATION = GR-QC/9610048;%%
\bibitem [{\citenamefont {Cardoso}(2004)}]{Cardoso:2004hh}%
  \BibitemOpen
  \bibfield  {author} {\bibinfo {author} {\bibfnamefont {V.}~\bibnamefont
  {Cardoso}},\ }\href {\doibase 10.1103/PhysRevD.70.127502} {\bibfield
  {journal} {\bibinfo  {journal} {Phys. Rev.}\ }\textbf {\bibinfo {volume}
  {D70}},\ \bibinfo {pages} {127502} (\bibinfo {year} {2004})},\ \Eprint
  {http://arxiv.org/abs/gr-qc/0411048} {arXiv:gr-qc/0411048 [gr-qc]}
  \BibitemShut {NoStop}%
%%CITATION = GR-QC/0411048;%%
\bibitem [{\citenamefont {Hod}(2008)}]{Hod:2008zz}%
  \BibitemOpen
  \bibfield  {author} {\bibinfo {author} {\bibfnamefont {S.}~\bibnamefont
  {Hod}},\ }\href {\doibase 10.1103/PhysRevD.78.084035} {\bibfield  {journal}
  {\bibinfo  {journal} {Phys. Rev.}\ }\textbf {\bibinfo {volume} {D78}},\
  \bibinfo {pages} {084035} (\bibinfo {year} {2008})},\ \Eprint
  {http://arxiv.org/abs/0811.3806} {arXiv:0811.3806 [gr-qc]} \BibitemShut
  {NoStop}%
%%CITATION = ARXIV:0811.3806;%%
\bibitem [{\citenamefont {Yang}\ \emph
  {et~al.}(2013{\natexlab{a}})\citenamefont {Yang}, \citenamefont {Zhang},
  \citenamefont {Zimmerman}, \citenamefont {Nichols}, \citenamefont {Berti},\
  and\ \citenamefont {Chen}}]{Yang:2012pj}%
  \BibitemOpen
  \bibfield  {author} {\bibinfo {author} {\bibfnamefont {H.}~\bibnamefont
  {Yang}}, \bibinfo {author} {\bibfnamefont {F.}~\bibnamefont {Zhang}},
  \bibinfo {author} {\bibfnamefont {A.}~\bibnamefont {Zimmerman}}, \bibinfo
  {author} {\bibfnamefont {D.~A.}\ \bibnamefont {Nichols}}, \bibinfo {author}
  {\bibfnamefont {E.}~\bibnamefont {Berti}}, \ and\ \bibinfo {author}
  {\bibfnamefont {Y.}~\bibnamefont {Chen}},\ }\href {\doibase
  10.1103/PhysRevD.87.041502} {\bibfield  {journal} {\bibinfo  {journal} {Phys.
  Rev.}\ }\textbf {\bibinfo {volume} {D87}},\ \bibinfo {pages} {041502}
  (\bibinfo {year} {2013}{\natexlab{a}})},\ \Eprint
  {http://arxiv.org/abs/1212.3271} {arXiv:1212.3271 [gr-qc]} \BibitemShut
  {NoStop}%
%%CITATION = ARXIV:1212.3271;%%
\bibitem [{\citenamefont {Yang}\ \emph
  {et~al.}(2013{\natexlab{b}})\citenamefont {Yang}, \citenamefont {Zimmerman},
  \citenamefont {Zenginoğlu}, \citenamefont {Zhang}, \citenamefont {Berti},\
  and\ \citenamefont {Chen}}]{Yang:2013uba}%
  \BibitemOpen
  \bibfield  {author} {\bibinfo {author} {\bibfnamefont {H.}~\bibnamefont
  {Yang}}, \bibinfo {author} {\bibfnamefont {A.}~\bibnamefont {Zimmerman}},
  \bibinfo {author} {\bibfnamefont {A.}~\bibnamefont {Zenginoğlu}}, \bibinfo
  {author} {\bibfnamefont {F.}~\bibnamefont {Zhang}}, \bibinfo {author}
  {\bibfnamefont {E.}~\bibnamefont {Berti}}, \ and\ \bibinfo {author}
  {\bibfnamefont {Y.}~\bibnamefont {Chen}},\ }\href {\doibase
  10.1103/PhysRevD.88.044047} {\bibfield  {journal} {\bibinfo  {journal} {Phys.
  Rev.}\ }\textbf {\bibinfo {volume} {D88}},\ \bibinfo {pages} {044047}
  (\bibinfo {year} {2013}{\natexlab{b}})},\ \bibinfo {note} {[Phys.
  Rev.D88,044047(2013)]},\ \Eprint {http://arxiv.org/abs/1307.8086}
  {arXiv:1307.8086 [gr-qc]} \BibitemShut {NoStop}%
%%CITATION = ARXIV:1307.8086;%%
\bibitem [{\citenamefont {Nakano}\ \emph {et~al.}(2016)\citenamefont {Nakano},
  \citenamefont {Sago}, \citenamefont {Tanaka},\ and\ \citenamefont
  {Nakamura}}]{Nakano:2016zvv}%
  \BibitemOpen
  \bibfield  {author} {\bibinfo {author} {\bibfnamefont {H.}~\bibnamefont
  {Nakano}}, \bibinfo {author} {\bibfnamefont {N.}~\bibnamefont {Sago}},
  \bibinfo {author} {\bibfnamefont {T.}~\bibnamefont {Tanaka}}, \ and\ \bibinfo
  {author} {\bibfnamefont {T.}~\bibnamefont {Nakamura}},\ }\href {\doibase
  10.1093/ptep/ptw098} {\bibfield  {journal} {\bibinfo  {journal} {PTEP}\
  }\textbf {\bibinfo {volume} {2016}},\ \bibinfo {pages} {083E01} (\bibinfo
  {year} {2016})},\ \Eprint {http://arxiv.org/abs/1604.08285} {arXiv:1604.08285
  [gr-qc]} \BibitemShut {NoStop}%
%%CITATION = ARXIV:1604.08285;%%
\bibitem [{\citenamefont {Gralla}\ \emph {et~al.}(2016)\citenamefont {Gralla},
  \citenamefont {Zimmerman},\ and\ \citenamefont {Zimmerman}}]{Gralla:2016sxp}%
  \BibitemOpen
  \bibfield  {author} {\bibinfo {author} {\bibfnamefont {S.~E.}\ \bibnamefont
  {Gralla}}, \bibinfo {author} {\bibfnamefont {A.}~\bibnamefont {Zimmerman}}, \
  and\ \bibinfo {author} {\bibfnamefont {P.}~\bibnamefont {Zimmerman}},\ }\href
  {\doibase 10.1103/PhysRevD.94.084017} {\bibfield  {journal} {\bibinfo
  {journal} {Phys. Rev.}\ }\textbf {\bibinfo {volume} {D94}},\ \bibinfo {pages}
  {084017} (\bibinfo {year} {2016})},\ \Eprint
  {http://arxiv.org/abs/1608.04739} {arXiv:1608.04739 [gr-qc]} \BibitemShut
  {NoStop}%
%%CITATION = ARXIV:1608.04739;%%
\bibitem [{\citenamefont {Zimmerman}(2016)}]{Zimmerman:2016qtn}%
  \BibitemOpen
  \bibfield  {author} {\bibinfo {author} {\bibfnamefont {P.}~\bibnamefont
  {Zimmerman}},\ }\href@noop {} {\  (\bibinfo {year} {2016})},\ \Eprint
  {http://arxiv.org/abs/1612.03172} {arXiv:1612.03172 [gr-qc]} \BibitemShut
  {NoStop}%
%%CITATION = ARXIV:1612.03172;%%
\bibitem [{\citenamefont {{Teukolsky}}(1973)}]{1973ApJ...185..635T}%
  \BibitemOpen
  \bibfield  {author} {\bibinfo {author} {\bibfnamefont {S.~A.}\ \bibnamefont
  {{Teukolsky}}},\ }\href {\doibase 10.1086/152444} {\bibfield  {journal}
  {\bibinfo  {journal} {\apj}\ }\textbf {\bibinfo {volume} {185}},\ \bibinfo
  {pages} {635} (\bibinfo {year} {1973})}\BibitemShut {NoStop}%
\bibitem [{\citenamefont {Berti}\ \emph {et~al.}(2006)\citenamefont {Berti},
  \citenamefont {Cardoso},\ and\ \citenamefont {Casals}}]{Berti:2005gp}%
  \BibitemOpen
  \bibfield  {author} {\bibinfo {author} {\bibfnamefont {E.}~\bibnamefont
  {Berti}}, \bibinfo {author} {\bibfnamefont {V.}~\bibnamefont {Cardoso}}, \
  and\ \bibinfo {author} {\bibfnamefont {M.}~\bibnamefont {Casals}},\ }\href
  {\doibase 10.1103/PhysRevD.73.109902, 10.1103/PhysRevD.73.024013} {\bibfield
  {journal} {\bibinfo  {journal} {Phys. Rev.}\ }\textbf {\bibinfo {volume}
  {D73}},\ \bibinfo {pages} {024013} (\bibinfo {year} {2006})},\ \bibinfo
  {note} {[Erratum: Phys. Rev.D73,109902(2006)]},\ \Eprint
  {http://arxiv.org/abs/gr-qc/0511111} {arXiv:gr-qc/0511111 [gr-qc]}
  \BibitemShut {NoStop}%
%%CITATION = GR-QC/0511111;%%
\bibitem [{\citenamefont {{Newman}}\ and\ \citenamefont
  {{Penrose}}(1962)}]{1962JMP.....3..566N}%
  \BibitemOpen
  \bibfield  {author} {\bibinfo {author} {\bibfnamefont {E.}~\bibnamefont
  {{Newman}}}\ and\ \bibinfo {author} {\bibfnamefont {R.}~\bibnamefont
  {{Penrose}}},\ }\href {\doibase 10.1063/1.1724257} {\bibfield  {journal}
  {\bibinfo  {journal} {Journal of Mathematical Physics}\ }\textbf {\bibinfo
  {volume} {3}},\ \bibinfo {pages} {566} (\bibinfo {year} {1962})}\BibitemShut
  {NoStop}%
\bibitem [{\citenamefont {{Kinnersley}}(1969)}]{1969JMP....10.1195K}%
  \BibitemOpen
  \bibfield  {author} {\bibinfo {author} {\bibfnamefont {W.}~\bibnamefont
  {{Kinnersley}}},\ }\href {\doibase 10.1063/1.1664958} {\bibfield  {journal}
  {\bibinfo  {journal} {Journal of Mathematical Physics}\ }\textbf {\bibinfo
  {volume} {10}},\ \bibinfo {pages} {1195} (\bibinfo {year}
  {1969})}\BibitemShut {NoStop}%
\bibitem [{\citenamefont {Teukolsky}(1972)}]{Teukolsky:1972my}%
  \BibitemOpen
  \bibfield  {author} {\bibinfo {author} {\bibfnamefont {S.~A.}\ \bibnamefont
  {Teukolsky}},\ }\href {\doibase 10.1103/PhysRevLett.29.1114} {\bibfield
  {journal} {\bibinfo  {journal} {Phys. Rev. Lett.}\ }\textbf {\bibinfo
  {volume} {29}},\ \bibinfo {pages} {1114} (\bibinfo {year}
  {1972})}\BibitemShut {NoStop}%
%%CITATION = PRLTA,29,1114;%%
\bibitem [{\citenamefont {Boyer}\ and\ \citenamefont
  {Lindquist}(1967)}]{Boyer:1966qh}%
  \BibitemOpen
  \bibfield  {author} {\bibinfo {author} {\bibfnamefont {R.~H.}\ \bibnamefont
  {Boyer}}\ and\ \bibinfo {author} {\bibfnamefont {R.~W.}\ \bibnamefont
  {Lindquist}},\ }\href {\doibase 10.1063/1.1705193} {\bibfield  {journal}
  {\bibinfo  {journal} {J. Math. Phys.}\ }\textbf {\bibinfo {volume} {8}},\
  \bibinfo {pages} {265} (\bibinfo {year} {1967})}\BibitemShut {NoStop}%
%%CITATION = JMAPA,8,265;%%
\bibitem [{\citenamefont {{Marcilhacy}}(1983)}]{1983NCimL..37..300M}%
  \BibitemOpen
  \bibfield  {author} {\bibinfo {author} {\bibfnamefont {G.}~\bibnamefont
  {{Marcilhacy}}},\ }\href@noop {} {\bibfield  {journal} {\bibinfo  {journal}
  {Nuovo Cimento Lettere}\ }\textbf {\bibinfo {volume} {37}},\ \bibinfo {pages}
  {300} (\bibinfo {year} {1983})}\BibitemShut {NoStop}%
\bibitem [{\citenamefont {{Blandin}}\ \emph {et~al.}(1983)\citenamefont
  {{Blandin}}, \citenamefont {{Pons}},\ and\ \citenamefont
  {{Marcilhacy}}}]{1983NCimL..38..561B}%
  \BibitemOpen
  \bibfield  {author} {\bibinfo {author} {\bibfnamefont {J.}~\bibnamefont
  {{Blandin}}}, \bibinfo {author} {\bibfnamefont {R.}~\bibnamefont {{Pons}}}, \
  and\ \bibinfo {author} {\bibfnamefont {G.}~\bibnamefont {{Marcilhacy}}},\
  }\href@noop {} {\bibfield  {journal} {\bibinfo  {journal} {Nuovo Cimento
  Lettere}\ }\textbf {\bibinfo {volume} {38}},\ \bibinfo {pages} {561}
  (\bibinfo {year} {1983})}\BibitemShut {NoStop}%
\bibitem [{\citenamefont {Fiziev}(2010)}]{Fiziev:2009wn}%
  \BibitemOpen
  \bibfield  {author} {\bibinfo {author} {\bibfnamefont {P.~P.}\ \bibnamefont
  {Fiziev}},\ }\href {\doibase 10.1088/0264-9381/27/13/135001} {\bibfield
  {journal} {\bibinfo  {journal} {Class. Quant. Grav.}\ }\textbf {\bibinfo
  {volume} {27}},\ \bibinfo {pages} {135001} (\bibinfo {year} {2010})},\
  \Eprint {http://arxiv.org/abs/0908.4234} {arXiv:0908.4234 [gr-qc]}
  \BibitemShut {NoStop}%
%%CITATION = ARXIV:0908.4234;%%
\bibitem [{\citenamefont {Cook}\ and\ \citenamefont
  {Zalutskiy}(2014)}]{Cook:2014cta}%
  \BibitemOpen
  \bibfield  {author} {\bibinfo {author} {\bibfnamefont {G.~B.}\ \bibnamefont
  {Cook}}\ and\ \bibinfo {author} {\bibfnamefont {M.}~\bibnamefont
  {Zalutskiy}},\ }\href {\doibase 10.1103/PhysRevD.90.124021} {\bibfield
  {journal} {\bibinfo  {journal} {Phys. Rev.}\ }\textbf {\bibinfo {volume}
  {D90}},\ \bibinfo {pages} {124021} (\bibinfo {year} {2014})},\ \Eprint
  {http://arxiv.org/abs/1410.7698} {arXiv:1410.7698 [gr-qc]} \BibitemShut
  {NoStop}%
%%CITATION = ARXIV:1410.7698;%%
\bibitem [{\citenamefont {{Zel'Dovich}}(1971)}]{1971JETPL..14..180Z}%
  \BibitemOpen
  \bibfield  {author} {\bibinfo {author} {\bibfnamefont {Y.~B.}\ \bibnamefont
  {{Zel'Dovich}}},\ }\href@noop {} {\bibfield  {journal} {\bibinfo  {journal}
  {Soviet Journal of Experimental and Theoretical Physics Letters}\ }\textbf
  {\bibinfo {volume} {14}},\ \bibinfo {pages} {180} (\bibinfo {year}
  {1971})}\BibitemShut {NoStop}%
\bibitem [{\citenamefont {{Zel'Dovich}}(1972)}]{1972JETP...35.1085Z}%
  \BibitemOpen
  \bibfield  {author} {\bibinfo {author} {\bibfnamefont {Y.~B.}\ \bibnamefont
  {{Zel'Dovich}}},\ }\href@noop {} {\bibfield  {journal} {\bibinfo  {journal}
  {Soviet Journal of Experimental and Theoretical Physics}\ }\textbf {\bibinfo
  {volume} {35}},\ \bibinfo {pages} {1085} (\bibinfo {year}
  {1972})}\BibitemShut {NoStop}%
\bibitem [{\citenamefont {{Misner}}(1972)}]{1972BAPS...17..472M}%
  \BibitemOpen
  \bibfield  {author} {\bibinfo {author} {\bibfnamefont {C.}~\bibnamefont
  {{Misner}}},\ }\href@noop {} {\bibfield  {journal} {\bibinfo  {journal}
  {Bulletin of the American Physical Society}\ }\textbf {\bibinfo {volume}
  {17}},\ \bibinfo {pages} {472} (\bibinfo {year} {1972})}\BibitemShut
  {NoStop}%
\bibitem [{\citenamefont {Starobinsky}(1973)}]{Starobinsky:1973aij}%
  \BibitemOpen
  \bibfield  {author} {\bibinfo {author} {\bibfnamefont {A.~A.}\ \bibnamefont
  {Starobinsky}},\ }\href@noop {} {\bibfield  {journal} {\bibinfo  {journal}
  {Sov. Phys. JETP}\ }\textbf {\bibinfo {volume} {37}},\ \bibinfo {pages} {28}
  (\bibinfo {year} {1973})},\ \bibinfo {note} {[Zh. Eksp. Teor.
  Fiz.64,48(1973)]}\BibitemShut {NoStop}%
%%CITATION = SPHJA,37,28;%%
\bibitem [{\citenamefont {{Starobinski{\v i}}}\ and\ \citenamefont
  {{Churilov}}(1974)}]{1974JETP...38....1S}%
  \BibitemOpen
  \bibfield  {author} {\bibinfo {author} {\bibfnamefont {A.~A.}\ \bibnamefont
  {{Starobinski{\v i}}}}\ and\ \bibinfo {author} {\bibfnamefont {S.~M.}\
  \bibnamefont {{Churilov}}},\ }\href@noop {} {\bibfield  {journal} {\bibinfo
  {journal} {Soviet Journal of Experimental and Theoretical Physics}\ }\textbf
  {\bibinfo {volume} {38}},\ \bibinfo {pages} {1} (\bibinfo {year}
  {1974})}\BibitemShut {NoStop}%
\bibitem [{\citenamefont {Bekenstein}\ and\ \citenamefont
  {Schiffer}(1998)}]{Bekenstein:1998nt}%
  \BibitemOpen
  \bibfield  {author} {\bibinfo {author} {\bibfnamefont {J.~D.}\ \bibnamefont
  {Bekenstein}}\ and\ \bibinfo {author} {\bibfnamefont {M.}~\bibnamefont
  {Schiffer}},\ }\href {\doibase 10.1103/PhysRevD.58.064014} {\bibfield
  {journal} {\bibinfo  {journal} {Phys. Rev.}\ }\textbf {\bibinfo {volume}
  {D58}},\ \bibinfo {pages} {064014} (\bibinfo {year} {1998})},\ \Eprint
  {http://arxiv.org/abs/gr-qc/9803033} {arXiv:gr-qc/9803033 [gr-qc]}
  \BibitemShut {NoStop}%
%%CITATION = GR-QC/9803033;%%
\bibitem [{\citenamefont {Richartz}\ \emph {et~al.}(2009)\citenamefont
  {Richartz}, \citenamefont {Weinfurtner}, \citenamefont {Penner},\ and\
  \citenamefont {Unruh}}]{Richartz:2009mi}%
  \BibitemOpen
  \bibfield  {author} {\bibinfo {author} {\bibfnamefont {M.}~\bibnamefont
  {Richartz}}, \bibinfo {author} {\bibfnamefont {S.}~\bibnamefont
  {Weinfurtner}}, \bibinfo {author} {\bibfnamefont {A.~J.}\ \bibnamefont
  {Penner}}, \ and\ \bibinfo {author} {\bibfnamefont {W.~G.}\ \bibnamefont
  {Unruh}},\ }\href {\doibase 10.1103/PhysRevD.80.124016} {\bibfield  {journal}
  {\bibinfo  {journal} {Phys. Rev.}\ }\textbf {\bibinfo {volume} {D80}},\
  \bibinfo {pages} {124016} (\bibinfo {year} {2009})},\ \Eprint
  {http://arxiv.org/abs/0909.2317} {arXiv:0909.2317 [gr-qc]} \BibitemShut
  {NoStop}%
%%CITATION = ARXIV:0909.2317;%%
\bibitem [{\citenamefont {Brito}\ \emph {et~al.}(2015)\citenamefont {Brito},
  \citenamefont {Cardoso},\ and\ \citenamefont {Pani}}]{Brito:2015oca}%
  \BibitemOpen
  \bibfield  {author} {\bibinfo {author} {\bibfnamefont {R.}~\bibnamefont
  {Brito}}, \bibinfo {author} {\bibfnamefont {V.}~\bibnamefont {Cardoso}}, \
  and\ \bibinfo {author} {\bibfnamefont {P.}~\bibnamefont {Pani}},\ }\href
  {\doibase 10.1007/978-3-319-19000-6} {\bibfield  {journal} {\bibinfo
  {journal} {Lect. Notes Phys.}\ }\textbf {\bibinfo {volume} {906}},\ \bibinfo
  {pages} {pp.1} (\bibinfo {year} {2015})},\ \Eprint
  {http://arxiv.org/abs/1501.06570} {arXiv:1501.06570 [gr-qc]} \BibitemShut
  {NoStop}%
%%CITATION = ARXIV:1501.06570;%%
\bibitem [{\citenamefont {Torres}\ \emph {et~al.}(2016)\citenamefont {Torres},
  \citenamefont {Patrick}, \citenamefont {Coutant}, \citenamefont {Richartz},
  \citenamefont {Tedford},\ and\ \citenamefont {Weinfurtner}}]{Torres:2016iee}%
  \BibitemOpen
  \bibfield  {author} {\bibinfo {author} {\bibfnamefont {T.}~\bibnamefont
  {Torres}}, \bibinfo {author} {\bibfnamefont {S.}~\bibnamefont {Patrick}},
  \bibinfo {author} {\bibfnamefont {A.}~\bibnamefont {Coutant}}, \bibinfo
  {author} {\bibfnamefont {M.}~\bibnamefont {Richartz}}, \bibinfo {author}
  {\bibfnamefont {E.~W.}\ \bibnamefont {Tedford}}, \ and\ \bibinfo {author}
  {\bibfnamefont {S.}~\bibnamefont {Weinfurtner}},\ }\href@noop {} {\
  (\bibinfo {year} {2016})},\ \Eprint {http://arxiv.org/abs/1612.06180}
  {arXiv:1612.06180 [gr-qc]} \BibitemShut {NoStop}%
%%CITATION = ARXIV:1612.06180;%%
\bibitem [{\citenamefont {Hod}(2012)}]{Hod:2012px}%
  \BibitemOpen
  \bibfield  {author} {\bibinfo {author} {\bibfnamefont {S.}~\bibnamefont
  {Hod}},\ }\href {\doibase 10.1103/PhysRevD.86.129902,
  10.1103/PhysRevD.86.104026} {\bibfield  {journal} {\bibinfo  {journal} {Phys.
  Rev.}\ }\textbf {\bibinfo {volume} {D86}},\ \bibinfo {pages} {104026}
  (\bibinfo {year} {2012})},\ \bibinfo {note} {[Erratum: Phys.
  Rev.D86,129902(2012)]},\ \Eprint {http://arxiv.org/abs/1211.3202}
  {arXiv:1211.3202 [gr-qc]} \BibitemShut {NoStop}%
%%CITATION = ARXIV:1211.3202;%%
\bibitem [{\citenamefont {Herdeiro}\ and\ \citenamefont
  {Radu}(2014)}]{Herdeiro:2014goa}%
  \BibitemOpen
  \bibfield  {author} {\bibinfo {author} {\bibfnamefont {C.~A.~R.}\
  \bibnamefont {Herdeiro}}\ and\ \bibinfo {author} {\bibfnamefont
  {E.}~\bibnamefont {Radu}},\ }\href {\doibase 10.1103/PhysRevLett.112.221101}
  {\bibfield  {journal} {\bibinfo  {journal} {Phys. Rev. Lett.}\ }\textbf
  {\bibinfo {volume} {112}},\ \bibinfo {pages} {221101} (\bibinfo {year}
  {2014})},\ \Eprint {http://arxiv.org/abs/1403.2757} {arXiv:1403.2757 [gr-qc]}
  \BibitemShut {NoStop}%
%%CITATION = ARXIV:1403.2757;%%
\bibitem [{\citenamefont {Herdeiro}\ and\ \citenamefont
  {Radu}(2015)}]{Herdeiro:2015gia}%
  \BibitemOpen
  \bibfield  {author} {\bibinfo {author} {\bibfnamefont {C.}~\bibnamefont
  {Herdeiro}}\ and\ \bibinfo {author} {\bibfnamefont {E.}~\bibnamefont
  {Radu}},\ }\href {\doibase 10.1088/0264-9381/32/14/144001} {\bibfield
  {journal} {\bibinfo  {journal} {Class. Quant. Grav.}\ }\textbf {\bibinfo
  {volume} {32}},\ \bibinfo {pages} {144001} (\bibinfo {year} {2015})},\
  \Eprint {http://arxiv.org/abs/1501.04319} {arXiv:1501.04319 [gr-qc]}
  \BibitemShut {NoStop}%
%%CITATION = ARXIV:1501.04319;%%
\bibitem [{\citenamefont {Hawking}\ \emph {et~al.}(1995)\citenamefont
  {Hawking}, \citenamefont {Horowitz},\ and\ \citenamefont
  {Ross}}]{Hawking:1994ii}%
  \BibitemOpen
  \bibfield  {author} {\bibinfo {author} {\bibfnamefont {S.~W.}\ \bibnamefont
  {Hawking}}, \bibinfo {author} {\bibfnamefont {G.~T.}\ \bibnamefont
  {Horowitz}}, \ and\ \bibinfo {author} {\bibfnamefont {S.~F.}\ \bibnamefont
  {Ross}},\ }\href {\doibase 10.1103/PhysRevD.51.4302} {\bibfield  {journal}
  {\bibinfo  {journal} {Phys. Rev.}\ }\textbf {\bibinfo {volume} {D51}},\
  \bibinfo {pages} {4302} (\bibinfo {year} {1995})},\ \Eprint
  {http://arxiv.org/abs/gr-qc/9409013} {arXiv:gr-qc/9409013 [gr-qc]}
  \BibitemShut {NoStop}%
%%CITATION = GR-QC/9409013;%%
\bibitem [{\citenamefont {Teitelboim}(1995)}]{Teitelboim:1994az}%
  \BibitemOpen
  \bibfield  {author} {\bibinfo {author} {\bibfnamefont {C.}~\bibnamefont
  {Teitelboim}},\ }\href {\doibase 10.1103/PhysRevD.52.6201,
  10.1103/PhysRevD.51.4315} {\bibfield  {journal} {\bibinfo  {journal} {Phys.
  Rev.}\ }\textbf {\bibinfo {volume} {D51}},\ \bibinfo {pages} {4315} (\bibinfo
  {year} {1995})},\ \bibinfo {note} {[Erratum: Phys. Rev.D52,6201(1995)]},\
  \Eprint {http://arxiv.org/abs/hep-th/9410103} {arXiv:hep-th/9410103 [hep-th]}
  \BibitemShut {NoStop}%
%%CITATION = HEP-TH/9410103;%%
\bibitem [{\citenamefont {Das}\ \emph {et~al.}(1997)\citenamefont {Das},
  \citenamefont {Dasgupta},\ and\ \citenamefont {Ramadevi}}]{Das:1996rn}%
  \BibitemOpen
  \bibfield  {author} {\bibinfo {author} {\bibfnamefont {S.}~\bibnamefont
  {Das}}, \bibinfo {author} {\bibfnamefont {A.}~\bibnamefont {Dasgupta}}, \
  and\ \bibinfo {author} {\bibfnamefont {P.}~\bibnamefont {Ramadevi}},\ }\href
  {\doibase 10.1142/S0217732397003186} {\bibfield  {journal} {\bibinfo
  {journal} {Mod. Phys. Lett.}\ }\textbf {\bibinfo {volume} {A12}},\ \bibinfo
  {pages} {3067} (\bibinfo {year} {1997})},\ \Eprint
  {http://arxiv.org/abs/hep-th/9608162} {arXiv:hep-th/9608162 [hep-th]}
  \BibitemShut {NoStop}%
%%CITATION = HEP-TH/9608162;%%
\bibitem [{\citenamefont {Carroll}\ \emph {et~al.}(2009)\citenamefont
  {Carroll}, \citenamefont {Johnson},\ and\ \citenamefont
  {Randall}}]{Carroll:2009maa}%
  \BibitemOpen
  \bibfield  {author} {\bibinfo {author} {\bibfnamefont {S.~M.}\ \bibnamefont
  {Carroll}}, \bibinfo {author} {\bibfnamefont {M.~C.}\ \bibnamefont
  {Johnson}}, \ and\ \bibinfo {author} {\bibfnamefont {L.}~\bibnamefont
  {Randall}},\ }\href {\doibase 10.1088/1126-6708/2009/11/109} {\bibfield
  {journal} {\bibinfo  {journal} {JHEP}\ }\textbf {\bibinfo {volume} {11}},\
  \bibinfo {pages} {109} (\bibinfo {year} {2009})},\ \Eprint
  {http://arxiv.org/abs/0901.0931} {arXiv:0901.0931 [hep-th]} \BibitemShut
  {NoStop}%
%%CITATION = ARXIV:0901.0931;%%
\bibitem [{\citenamefont {Pradhan}\ and\ \citenamefont
  {Majumdar}(2011)}]{Pradhan:2010ws}%
  \BibitemOpen
  \bibfield  {author} {\bibinfo {author} {\bibfnamefont {P.}~\bibnamefont
  {Pradhan}}\ and\ \bibinfo {author} {\bibfnamefont {P.}~\bibnamefont
  {Majumdar}},\ }\href {\doibase 10.1016/j.physleta.2010.11.015} {\bibfield
  {journal} {\bibinfo  {journal} {Phys. Lett.}\ }\textbf {\bibinfo {volume}
  {A375}},\ \bibinfo {pages} {474} (\bibinfo {year} {2011})},\ \Eprint
  {http://arxiv.org/abs/1001.0359} {arXiv:1001.0359 [gr-qc]} \BibitemShut
  {NoStop}%
%%CITATION = ARXIV:1001.0359;%%
\bibitem [{\citenamefont {Pradhan}\ and\ \citenamefont
  {Majumdar}(2013)}]{Pradhan:2012yx}%
  \BibitemOpen
  \bibfield  {author} {\bibinfo {author} {\bibfnamefont {P.}~\bibnamefont
  {Pradhan}}\ and\ \bibinfo {author} {\bibfnamefont {P.}~\bibnamefont
  {Majumdar}},\ }\href {\doibase 10.1140/epjc/s10052-013-2470-2} {\bibfield
  {journal} {\bibinfo  {journal} {Eur. Phys. J.}\ }\textbf {\bibinfo {volume}
  {C73}},\ \bibinfo {pages} {2470} (\bibinfo {year} {2013})},\ \Eprint
  {http://arxiv.org/abs/1108.2333} {arXiv:1108.2333 [gr-qc]} \BibitemShut
  {NoStop}%
%%CITATION = ARXIV:1108.2333;%%
\bibitem [{\citenamefont {Gralla}\ \emph {et~al.}(2015)\citenamefont {Gralla},
  \citenamefont {Porfyriadis},\ and\ \citenamefont
  {Warburton}}]{Gralla:2015rpa}%
  \BibitemOpen
  \bibfield  {author} {\bibinfo {author} {\bibfnamefont {S.~E.}\ \bibnamefont
  {Gralla}}, \bibinfo {author} {\bibfnamefont {A.~P.}\ \bibnamefont
  {Porfyriadis}}, \ and\ \bibinfo {author} {\bibfnamefont {N.}~\bibnamefont
  {Warburton}},\ }\href {\doibase 10.1103/PhysRevD.92.064029} {\bibfield
  {journal} {\bibinfo  {journal} {Phys. Rev.}\ }\textbf {\bibinfo {volume}
  {D92}},\ \bibinfo {pages} {064029} (\bibinfo {year} {2015})},\ \Eprint
  {http://arxiv.org/abs/1506.08496} {arXiv:1506.08496 [gr-qc]} \BibitemShut
  {NoStop}%
%%CITATION = ARXIV:1506.08496;%%
\bibitem [{\citenamefont {{Bardeen}}\ \emph {et~al.}(1972)\citenamefont
  {{Bardeen}}, \citenamefont {{Press}},\ and\ \citenamefont
  {{Teukolsky}}}]{1972ApJ...178..347B}%
  \BibitemOpen
  \bibfield  {author} {\bibinfo {author} {\bibfnamefont {J.~M.}\ \bibnamefont
  {{Bardeen}}}, \bibinfo {author} {\bibfnamefont {W.~H.}\ \bibnamefont
  {{Press}}}, \ and\ \bibinfo {author} {\bibfnamefont {S.~A.}\ \bibnamefont
  {{Teukolsky}}},\ }\href {\doibase 10.1086/151796} {\bibfield  {journal}
  {\bibinfo  {journal} {\apj}\ }\textbf {\bibinfo {volume} {178}},\ \bibinfo
  {pages} {347} (\bibinfo {year} {1972})}\BibitemShut {NoStop}%
\bibitem [{\citenamefont {{Hawking}}\ and\ \citenamefont
  {{Hartle}}(1972)}]{1972CMaPh..27..283H}%
  \BibitemOpen
  \bibfield  {author} {\bibinfo {author} {\bibfnamefont {S.~W.}\ \bibnamefont
  {{Hawking}}}\ and\ \bibinfo {author} {\bibfnamefont {J.~B.}\ \bibnamefont
  {{Hartle}}},\ }\href {\doibase 10.1007/BF01645515} {\bibfield  {journal}
  {\bibinfo  {journal} {Communications in Mathematical Physics}\ }\textbf
  {\bibinfo {volume} {27}},\ \bibinfo {pages} {283} (\bibinfo {year}
  {1972})}\BibitemShut {NoStop}%
\bibitem [{\citenamefont {Leaver}(1985)}]{Leaver:1985ax}%
  \BibitemOpen
  \bibfield  {author} {\bibinfo {author} {\bibfnamefont {E.~W.}\ \bibnamefont
  {Leaver}},\ }\href {\doibase 10.1098/rspa.1985.0119} {\bibfield  {journal}
  {\bibinfo  {journal} {Proc. Roy. Soc. Lond.}\ }\textbf {\bibinfo {volume}
  {A402}},\ \bibinfo {pages} {285} (\bibinfo {year} {1985})}\BibitemShut
  {NoStop}%
%%CITATION = PRSLA,A402,285;%%
\bibitem [{\citenamefont {Onozawa}\ \emph {et~al.}(1996)\citenamefont
  {Onozawa}, \citenamefont {Mishima}, \citenamefont {Okamura},\ and\
  \citenamefont {Ishihara}}]{Onozawa:1995vu}%
  \BibitemOpen
  \bibfield  {author} {\bibinfo {author} {\bibfnamefont {H.}~\bibnamefont
  {Onozawa}}, \bibinfo {author} {\bibfnamefont {T.}~\bibnamefont {Mishima}},
  \bibinfo {author} {\bibfnamefont {T.}~\bibnamefont {Okamura}}, \ and\
  \bibinfo {author} {\bibfnamefont {H.}~\bibnamefont {Ishihara}},\ }\href
  {\doibase 10.1103/PhysRevD.53.7033} {\bibfield  {journal} {\bibinfo
  {journal} {Phys. Rev.}\ }\textbf {\bibinfo {volume} {D53}},\ \bibinfo {pages}
  {7033} (\bibinfo {year} {1996})},\ \Eprint
  {http://arxiv.org/abs/gr-qc/9603021} {arXiv:gr-qc/9603021 [gr-qc]}
  \BibitemShut {NoStop}%
%%CITATION = GR-QC/9603021;%%
\bibitem [{\citenamefont {{Hod}}(2010)}]{2010PhLA..374.2901H}%
  \BibitemOpen
  \bibfield  {author} {\bibinfo {author} {\bibfnamefont {S.}~\bibnamefont
  {{Hod}}},\ }\href {\doibase 10.1016/j.physleta.2010.05.052} {\bibfield
  {journal} {\bibinfo  {journal} {Physics Letters A}\ }\textbf {\bibinfo
  {volume} {374}},\ \bibinfo {pages} {2901} (\bibinfo {year} {2010})},\ \Eprint
  {http://arxiv.org/abs/1006.4439} {arXiv:1006.4439 [gr-qc]} \BibitemShut
  {NoStop}%
\bibitem [{\citenamefont {Maassen van~den
  Brink}(2000)}]{MaassenvandenBrink:2000iwh}%
  \BibitemOpen
  \bibfield  {author} {\bibinfo {author} {\bibfnamefont {A.}~\bibnamefont
  {Maassen van~den Brink}},\ }\href {\doibase 10.1103/PhysRevD.62.064009}
  {\bibfield  {journal} {\bibinfo  {journal} {Phys. Rev.}\ }\textbf {\bibinfo
  {volume} {D62}},\ \bibinfo {pages} {064009} (\bibinfo {year} {2000})},\
  \Eprint {http://arxiv.org/abs/gr-qc/0001032} {arXiv:gr-qc/0001032 [gr-qc]}
  \BibitemShut {NoStop}%
%%CITATION = GR-QC/0001032;%%
\bibitem [{\citenamefont {Berti}\ \emph {et~al.}(2003)\citenamefont {Berti},
  \citenamefont {Cardoso}, \citenamefont {Kokkotas},\ and\ \citenamefont
  {Onozawa}}]{Berti:2003jh}%
  \BibitemOpen
  \bibfield  {author} {\bibinfo {author} {\bibfnamefont {E.}~\bibnamefont
  {Berti}}, \bibinfo {author} {\bibfnamefont {V.}~\bibnamefont {Cardoso}},
  \bibinfo {author} {\bibfnamefont {K.~D.}\ \bibnamefont {Kokkotas}}, \ and\
  \bibinfo {author} {\bibfnamefont {H.}~\bibnamefont {Onozawa}},\ }\href
  {\doibase 10.1103/PhysRevD.68.124018} {\bibfield  {journal} {\bibinfo
  {journal} {Phys. Rev.}\ }\textbf {\bibinfo {volume} {D68}},\ \bibinfo {pages}
  {124018} (\bibinfo {year} {2003})},\ \Eprint
  {http://arxiv.org/abs/hep-th/0307013} {arXiv:hep-th/0307013 [hep-th]}
  \BibitemShut {NoStop}%
%%CITATION = HEP-TH/0307013;%%
\bibitem [{\citenamefont {Benone}\ \emph {et~al.}(2014)\citenamefont {Benone},
  \citenamefont {Crispino}, \citenamefont {Herdeiro},\ and\ \citenamefont
  {Radu}}]{Benone:2014ssa}%
  \BibitemOpen
  \bibfield  {author} {\bibinfo {author} {\bibfnamefont {C.~L.}\ \bibnamefont
  {Benone}}, \bibinfo {author} {\bibfnamefont {L.~C.~B.}\ \bibnamefont
  {Crispino}}, \bibinfo {author} {\bibfnamefont {C.}~\bibnamefont {Herdeiro}},
  \ and\ \bibinfo {author} {\bibfnamefont {E.}~\bibnamefont {Radu}},\ }\href
  {\doibase 10.1103/PhysRevD.90.104024} {\bibfield  {journal} {\bibinfo
  {journal} {Phys. Rev.}\ }\textbf {\bibinfo {volume} {D90}},\ \bibinfo {pages}
  {104024} (\bibinfo {year} {2014})},\ \Eprint {http://arxiv.org/abs/1409.1593}
  {arXiv:1409.1593 [gr-qc]} \BibitemShut {NoStop}%
%%CITATION = ARXIV:1409.1593;%%
\bibitem [{\citenamefont {East}\ and\ \citenamefont
  {Pretorius}(2017)}]{East:2017ovw}%
  \BibitemOpen
  \bibfield  {author} {\bibinfo {author} {\bibfnamefont {W.~E.}\ \bibnamefont
  {East}}\ and\ \bibinfo {author} {\bibfnamefont {F.}~\bibnamefont
  {Pretorius}},\ }\href@noop {} {\  (\bibinfo {year} {2017})},\ \Eprint
  {http://arxiv.org/abs/1704.04791} {arXiv:1704.04791 [gr-qc]} \BibitemShut
  {NoStop}%
%%CITATION = ARXIV:1704.04791;%%
\bibitem [{\citenamefont {Unruh}(1973)}]{Unruh:1973bda}%
  \BibitemOpen
  \bibfield  {author} {\bibinfo {author} {\bibfnamefont {W.}~\bibnamefont
  {Unruh}},\ }\href {\doibase 10.1103/PhysRevLett.31.1265} {\bibfield
  {journal} {\bibinfo  {journal} {Phys. Rev. Lett.}\ }\textbf {\bibinfo
  {volume} {31}},\ \bibinfo {pages} {1265} (\bibinfo {year}
  {1973})}\BibitemShut {NoStop}%
%%CITATION = PRLTA,31,1265;%%
\bibitem [{\citenamefont {Torres~del Castillo}\ and\ \citenamefont
  {Silva-Ortigoza}(1990)}]{TorresdelCastillo:1990aw}%
  \BibitemOpen
  \bibfield  {author} {\bibinfo {author} {\bibfnamefont {G.~F.}\ \bibnamefont
  {Torres~del Castillo}}\ and\ \bibinfo {author} {\bibfnamefont
  {G.}~\bibnamefont {Silva-Ortigoza}},\ }\href {\doibase
  10.1103/PhysRevD.42.4082} {\bibfield  {journal} {\bibinfo  {journal} {Phys.
  Rev.}\ }\textbf {\bibinfo {volume} {D42}},\ \bibinfo {pages} {4082} (\bibinfo
  {year} {1990})}\BibitemShut {NoStop}%
%%CITATION = PHRVA,D42,4082;%%
\bibitem [{\citenamefont {Page}(1976)}]{Page:1976jj}%
  \BibitemOpen
  \bibfield  {author} {\bibinfo {author} {\bibfnamefont {D.~N.}\ \bibnamefont
  {Page}},\ }\href {\doibase 10.1103/PhysRevD.14.1509} {\bibfield  {journal}
  {\bibinfo  {journal} {Phys. Rev.}\ }\textbf {\bibinfo {volume} {D14}},\
  \bibinfo {pages} {1509} (\bibinfo {year} {1976})}\BibitemShut {NoStop}%
%%CITATION = PHRVA,D14,1509;%%
\bibitem [{\citenamefont {Gibbons}(1975)}]{Gibbons:1975kk}%
  \BibitemOpen
  \bibfield  {author} {\bibinfo {author} {\bibfnamefont {G.~W.}\ \bibnamefont
  {Gibbons}},\ }\href {\doibase 10.1007/BF01609829} {\bibfield  {journal}
  {\bibinfo  {journal} {Commun. Math. Phys.}\ }\textbf {\bibinfo {volume}
  {44}},\ \bibinfo {pages} {245} (\bibinfo {year} {1975})}\BibitemShut
  {NoStop}%
%%CITATION = CMPHA,44,245;%%
\bibitem [{\citenamefont {{Richartz}}\ and\ \citenamefont
  {{Saa}}(2011)}]{2011PhRvD..84j4021R}%
  \BibitemOpen
  \bibfield  {author} {\bibinfo {author} {\bibfnamefont {M.}~\bibnamefont
  {{Richartz}}}\ and\ \bibinfo {author} {\bibfnamefont {A.}~\bibnamefont
  {{Saa}}},\ }\href {\doibase 10.1103/PhysRevD.84.104021} {\bibfield  {journal}
  {\bibinfo  {journal} {\prd}\ }\textbf {\bibinfo {volume} {84}},\ \bibinfo
  {eid} {104021} (\bibinfo {year} {2011})},\ \Eprint
  {http://arxiv.org/abs/1109.3364} {arXiv:1109.3364 [gr-qc]} \BibitemShut
  {NoStop}%
\bibitem [{\citenamefont {Degollado}\ and\ \citenamefont
  {Herdeiro}(2013)}]{Degollado:2013eqa}%
  \BibitemOpen
  \bibfield  {author} {\bibinfo {author} {\bibfnamefont {J.~C.}\ \bibnamefont
  {Degollado}}\ and\ \bibinfo {author} {\bibfnamefont {C.~A.~R.}\ \bibnamefont
  {Herdeiro}},\ }\href {\doibase 10.1007/s10714-013-1598-6} {\bibfield
  {journal} {\bibinfo  {journal} {Gen. Rel. Grav.}\ }\textbf {\bibinfo {volume}
  {45}},\ \bibinfo {pages} {2483} (\bibinfo {year} {2013})},\ \Eprint
  {http://arxiv.org/abs/1303.2392} {arXiv:1303.2392 [gr-qc]} \BibitemShut
  {NoStop}%
%%CITATION = ARXIV:1303.2392;%%
\bibitem [{\citenamefont {Sampaio}\ \emph {et~al.}(2014)\citenamefont
  {Sampaio}, \citenamefont {Herdeiro},\ and\ \citenamefont
  {Wang}}]{Sampaio:2014swa}%
  \BibitemOpen
  \bibfield  {author} {\bibinfo {author} {\bibfnamefont {M.~O.~P.}\
  \bibnamefont {Sampaio}}, \bibinfo {author} {\bibfnamefont {C.}~\bibnamefont
  {Herdeiro}}, \ and\ \bibinfo {author} {\bibfnamefont {M.}~\bibnamefont
  {Wang}},\ }\href {\doibase 10.1103/PhysRevD.90.064004} {\bibfield  {journal}
  {\bibinfo  {journal} {Phys. Rev.}\ }\textbf {\bibinfo {volume} {D90}},\
  \bibinfo {pages} {064004} (\bibinfo {year} {2014})},\ \Eprint
  {http://arxiv.org/abs/1406.3536} {arXiv:1406.3536 [gr-qc]} \BibitemShut
  {NoStop}%
%%CITATION = ARXIV:1406.3536;%%
\bibitem [{\citenamefont {Richartz}\ and\ \citenamefont
  {Giugno}(2014)}]{Richartz:2014jla}%
  \BibitemOpen
  \bibfield  {author} {\bibinfo {author} {\bibfnamefont {M.}~\bibnamefont
  {Richartz}}\ and\ \bibinfo {author} {\bibfnamefont {D.}~\bibnamefont
  {Giugno}},\ }\href {\doibase 10.1103/PhysRevD.90.124011} {\bibfield
  {journal} {\bibinfo  {journal} {Phys. Rev.}\ }\textbf {\bibinfo {volume}
  {D90}},\ \bibinfo {pages} {124011} (\bibinfo {year} {2014})},\ \Eprint
  {http://arxiv.org/abs/1409.7440} {arXiv:1409.7440 [gr-qc]} \BibitemShut
  {NoStop}%
%%CITATION = ARXIV:1409.7440;%%
\end{thebibliography}%

\end{document}